\documentclass[]{aa} 

\usepackage{graphicx}
\usepackage{txfonts}
\usepackage{natbib}
\usepackage{subfigure}
\usepackage{float}
\begin{document}

\title{Image modeling of compact starburst clusters \thanks{The effect of primordial mass segregation, binary fraction and stellar evolution on the evolution of massive clusters on their early life-stages}: I. R136 }
\authorrunning{Khorrami et al.}  \titlerunning{Image modeling of compact starburst clusters: R136}

 \author{Z. Khorrami \inst{1}  \and
         F. Vakili \inst{1}  \and
         O. Chesneau \inst{1} 
} 
          
\offprints{zeinab.khorrami@oca.eu}

\institute{Laboratoire Lagrange, Universit\'{e} C\^{o}te d'Azur, Observatoire de la C\^{o}te d'Azur, CNRS,\\
Boulevard de l'Observatoire, F-06304 Nice, France\\
  \email{zeinab.khorrami@oca.eu}%
}

\date{}

 
  \abstract
  {Continuous progress in data quality from HST, recent multiwavelength high resolution spectroscopy and high contrast imaging from ground adaptive optics on large telescopes need exhaustive modeling of R136 to better understand its nature and evolutionary stage.}
  {To produce the best synthesized multiwavelength images of R136 we need to simulate the effect of dynamical and stellar evolution, mass segregation and binary stars fraction on the survival of young massive clusters with the initial parameters of R136 in the LMC, being set to the state of the art knowledge of this famous cluster.}
  {We produced a series of 32 young massive clusters using the NBODY6 code. Each cluster was tracked with adequate temporal samples to correctly follow the evolution of R136 during its early stages. 
To compare the results of the NBODY6 simulations with observational imaging data, we created the synthetic images from the output of the code.
We used the TLUSTY (for massive O and B stars) and KURUCZ (for other spectral types) model atmospheres to produce the fluxes in HST/ WFPC2 filters. GENEVA isochrones grids were used to track the evolution of stars.
Then, we derived the observable parameters from synthetic scenes at the spatial resolution of HST/WFPC2 in the F814W filter corresponding to 790.48nm central wavelength. Surface brightness profile of the cluster, half-light radius, mass function and neighbor radius were used as criteria to select the best representation of R136.
}
  {We compared the simulations of R136 to its HST imagery in recent years by creating synthetic scenes at the same resolution, pixel scale and FOV of the HST. We applied the same photometric analysis of the images as of the real ones. Having extracted the stellar sources, we estimated the mass-function (MF), the surface brightness profile (SBP), the half-light radius and the neighbor radius, $R_{neighbor}$ across R136. The interpretation of these four criteria point to the fact that an initially non-segregated cluster scenario is more representative of R136. This result pleads for the formation of massive stars by accretion rather than by collision.}
  {}

                
\maketitle


\section{Introduction}
R136 is known to be a very massive ($10^5 \rm{M}_{\odot}$), young star cluster at the center of 30 Doradus star forming region in the LMC. This cluster is remarkable for its large number of massive stars formed in a relatively small volume (\citet{hunter95}). 
The total number of O3 stars in this cluster, at least 41, exceeds the total number known elsewhere in the Milky Way or Magellanic Clouds (\citet{massey98}). 
The WN stars in R136 are presumably core H-burning stars that mimic WR-like spectra due to their very high intrinsic luminosity. Here the individual luminosities are a factor of 10 higher than those of normal WN stars of similar type, only encountered in the Galactic cluster NGC3603, that they also resemble spectroscopically (\citet{massey98}). 

R136 is outranking among other clusters in three aspects:

1) It is massive enough to produce such very massive stars as a relationship appears between the mass of a cluster and its highest-mass stars (\citet{weidner06}, \citet{weidner10}).

2) It is young enough for its massive members to be observed. The age of the cluster in the core is $2\pm1$Myr (\citet{deKoter98}, \citet{massey98}, \citet{crowther2010}).

3) It is close enough to be spatially resolved. We considered the distance of 48.5 kpc for our simulations which agrees well with \citet{Selman99} and \citet{gieren98}.

Therefore R136 can provide clues to understand the unsolved problems of formation and evolution of massive stars and star clusters, initial distribution of stellar masses at birth (IMF), if the massive stars tend to be formed locally in the center of the cloud (mass segregation) and the binary-multiple fraction of these stars as this property can impact the rate of SNe produced soon after a few Myr evolution of such clusters.

In the present study, unlike usual studies of R136 in the literature (\citet{banerjee2013}), we have adopted a new approach to unveil the different facets of R136. 
Our approach is primarily motivated by the fact that presently operated 8-10m class optical telescopes and foreseen extremely large telescopes like E-ELT should deliver diffraction limited visible and IR images of crowded field clusters such as R136.

Having this in mind we created a series of simulated multi-color images of R136 from the output of the numerical dynamical model (NBODY6 code, \cite{Aarseth2003}). In this work, we present the results from the comparison of the HST/WFPC2 imaging data with synthesized images  from the output of the NBODY6 simulations at the age of 2 Myr. 
For this we used Geneva stellar evolution models (\cite{geneva}) and TLUSTY (\cite{tlustyO}) or KURUCZ (\cite{kurucz97}) model atmospheres depending on the spectral type of stars to calculate their flux in WFPC2/F814W filter at the age of 2 Myr.

The present paper is organized as followings. In Section \ref{sec:modeling} we shortly describe the NBODY6 code with special emphasis on the options and parameters that are relevant to the present work. We outline the initial conditions from which the evolution of R136 can be simulated and recorded following different tracks. 
Section \ref{sec:simresults} describes the results of these simulations, including their observational aspects.
These results are compared to the HST observations 
\footnote{Based on observations made with the NASA/ESA Hubble Space Telescope, obtained from the data archive at the Space Telescope Institute. STScI is operated by the association of Universities for Research in Astronomy, Inc. under the NASA contract NAS 5-26555.}
in Section \ref{sec:compareobs}. We introduce our method to create the synthetic images from the output of the NBODY6 code. For this comparison we define a number of criteria applied to simulated scenes from HST versus observed scenes. 
The relevance of the comparisons is then discussed in the Section \ref{sec:discussion}. While we derive some general and preliminary results in perspective of high dynamic, high dynamic and spatially resolved images of R136 to become soon available with SPHERE and GPI in the coming years in Section \ref{sec:conclusion} .


\section{Modeling a young massive star cluster}\label{sec:modeling}
We used NBODY6\footnote{http://www.ast.cam.ac.uk/~sverre/web/pages/nbody.htm} code which includes the individual stellar equations of motion for all members of the cluster without any simplifying assumption and approximation (\citet{Aarseth2003}). We remind that NBODY6 integrates the particle orbits using the highly accurate fourth-order Hermite scheme and deals with the diverging gravitational forces in close encounters through regularizations. In addition, this code can track the evolution of the individual stars since it employs the well-tested Single Star Evolution (SSE) and Binary Star Evolution (BSE) recipes (\citet{hurley2000} and \citet{hurley2002}).

The whole modeling intends to better understand the nature of R136 as it can be imaged nowadays, by taking into account the parameters and different mechanisms  that drive the evolution of the cluster such as the degree of initial mass segregation, initial binary fraction, lower/upper stellar masses and stellar evolution. The initial parameters for R136 have been set to the values that represent the best this cluster according to its general properties.

\subsection{Physical conditions and Mechanisms}\label{sec:physical}
Stellar clusters are born embedded in giant molecular clouds, with a few percent of them surviving and becoming bound clusters (\citet{lada2003}). It is generally admitted that the fate of the clusters must occur during the early stages of their evolution. In massive star-burst clusters, such as R136,  dynamical evolution of the cluster can be affected by their significant number of massive stars. The initial distribution of these massive stars (mass segregation) in space, specially if they form in binary systems, plays an important role on the evolution of the cluster. 

In order to include such effects, we simulated full segregated clusters (in which most massive stars are located deeper in the core) versus non-segregated clusters (in which the massive stars are randomly distributed in space). The result of these simulations can tell us if R136 is more similar to an initially segregated cluster or not. A result that should put important constraints on theories of massive star formation and the cluster consequent evolution as a whole.

Following clusters without initial binaries, we also considered clusters with 30 and 60 \% initial binaries which is not far from the observation as lower-limit of the spectroscopic binaries detected in the clusters of 30 Dor is found to be 45\% (\citet{bosch2009}). 

Finally, by comparing the results from clusters with and without stellar evolution, one can check for the effect of stellar evolution on the binaries and the dynamical effect of the binaries themselves on the evolution of the cluster.  

\subsection{Initial conditions}\label{sec:initial}
Initial setup of the clusters made by MCLUSTER \footnote{https://ahwkuepper.wordpress.com/mcluster/}(\citet{kupper11}).
The simulation time is limited to an age of 4 Myr with 0.1Myr time-steps. The total mass of the cluster is estimated (Selman \& Melnick 2013) to be in the range of:

$4.6 \times 10^4 M_{\odot} < M_{\odot} < 1.3 \times 10^5 M_{\odot}$ 

Therefore we adopt the total mass of the cluster to be $1.0 \times 10^5 M_{\odot}$. Kroupa IMF with mass ranging between $0.2M_{\odot}$ or $0.5M_{\odot}$  to $150 M_{\odot}$. For the density profile we adopted Plummer model (\citet{aarseth1974}). Simulated clusters are assumed isolated and not impacted by tidal fields (tf=0). The initial half mass radius of the clusters is 0.8 pc. And finally the metallicity of LMC (R136) is taken as half of the solar metallicity.
All the clusters are in virial equilibrium  as \citet{henault12} find that R136 is in virial equilibrium.

The set-up of the numerical experimentations is as follows: half of the clusters are segregated (s=1.0) and the others are non-segregated (s=0.0). We adopted 60\%, 30\% and 0\% percent of initial binary  for the simulated clusters.
There will be two groups of clusters. These groups are totally similar to each other (even the initial position and velocity of stars) but for one group stellar evolution is ON and for other group stellar evolution is OFF to see the effect of stellar evolution (Table~\ref{table:sims}). 

Table~\ref{table:sims} depicts the main characteristics of simulated clusters. 
The first column is the name of the cluster. The second column says if stellar evolution is ON or OFF. Column 3 corresponds to the degree of mass segregation, 0.0 means non-segregation and 1.0 means the cluster is fully segregated. The fourth column corresponds to the binary fraction.  The fifth column defines the number of initial binaries. $M_{min}$ and $M_{max}$ are the lower and upper mass cutoff respectively of initial mass function (The canonical Kroupa, \citet{kroupa2001a} ) and N is the initial number of stars. 

For the initial number of binaries we used random pairing but separate pairing for components with $m > m_{sort}$. In this way, pairing of primary and secondary components of binary stars above $m_{sort}$ are randomly paired among each other. The motivation for this lies in extensive observational data showing that massive O, B stars are more likely  to be found in an equal mass binary system (\cite{sana2011}). $m_{sort}$ is equal to $5 M_{\odot}$ in agreement with \cite{Kobulnicky2007}. More detail about semi-major axis and period distribution of binaries can be found in section \ref{sec:binary}.

The period distribution was taken from the \cite{kroupa1995a} period distribution since it unifies the observed Galactic field and pre-main sequence populations (see also \cite{kroupa2008}). But for massive binaries with $M_{primary} > m_{sort}$ we used the period distribution from Sana\&Evans 2011 since the period distribution of massive O,B spectroscopic binaries has been found to be significantly different from what is observed for low-mass binaries (\cite{sana2011}). Massive binaries are found to have short periods in the range from 2 days to 10 years with a peak at 10 days.

Eccentricities are assumed to have a thermal distribution i.e. f(e)=2e (\citet{kroupa2008}) and for the eccentricities of high mass binaries the distribution is taken according to \cite{sana2011}, leading to the computation of the semi-major axis (a) of each binary.

\begin{table*}
\centering 
\begin{tabular}{|c | c c c c c c c|} 
Model & SE & Seg & BF & $N_{bin}$& $M_{min}$ & $M_{max}$ & $N$\\
\hline 
\hline
S00seg00bin05m150 & ON& 0.0 & 0.0 & 0&0.5 &150 & 55992  \\
D00seg00bin05m150 & OFF& 0.0 & 0.0 & 0&0.5 &150 & 55992 \\
S10seg00bin05m150 & ON& 1.0 & 0.0 & 0&0.5 &150 & 55815 \\
D10seg00bin05m150 & OFF& 1.0 & 0.0 & 0&0.5 &150 & 55815 \\
\hline
S00seg03bin05m150 & ON& 0.0 & 0.3 & 8404&0.5 &150 & 56032 \\
D00seg03bin05m150 & OFF& 0.0 & 0.3 & 8404&0.5 &150 & 56032 \\
S10seg03bin05m150 & ON& 1.0 & 0.3 & 8536&0.5 &150 & 56908 \\
D10seg03bin05m150 & OFF& 1.0 & 0.3 & 8536&0.5 &150 & 56908 \\
\hline
S00seg06bin05m150 & ON& 0.0 & 0.6 & 17000&0.5 &150 & 56669 \\
D00seg06bin05m150 & OFF& 0.0 & 0.6 & 17000&0.5 &150 & 56669 \\
S10seg06bin05m150 & ON& 1.0 & 0.6 & 16686&0.5 &150 & 55622 \\
D10seg06bin05m150 & OFF& 1.0 & 0.6 & 16686&0.5 &150 & 55622 \\
\hline
\hline
S00seg00bin02m150 & ON& 0.0 & 0.0 & 0&0.2 &150 & 105666 \\
D00seg00bin02m150 & OFF& 0.0 & 0.0 & 0&0.2 &150 & 105666 \\
S10seg00bin02m150 & ON& 1.0 & 0.0 & 0&0.2 &150 & 107722 \\
D10seg00bin02m150 & OFF& 1.0 & 0.0 & 0&0.2 &150 & 107722\\
\hline
S00seg03bin02m150 & ON & 0.0  & 0.3 & 16134&0.2 &150 & 107565 \\
D00seg03bin02m150 & OFF& 0.0 & 0.3 & 16134&0.2 &150 & 107565 \\
S10seg03bin02m150 & ON  & 1.0 & 0.3 & 16084&0.2 &150 & 107230 \\
D10seg03bin02m150 & OFF& 1.0 & 0.3 & 16084&0.2 &150 & 107230 \\
\hline
S00seg06bin02m150 & ON & 0.0  & 0.6 & 32225&0.2 &150 & 107419 \\
D00seg06bin02m150 & OFF& 0.0 & 0.6 & 32225&0.2 &150 & 107419 \\
S10seg06bin02m150 & ON  & 1.0 & 0.6 & 32309&0.2 &150 & 107698 \\
D10seg06bin02m150 & OFF& 1.0 & 0.6 & 32309&0.2 &150 & 107698 \\
\hline 
\end{tabular} 
\caption{Different simulated clusters grouped by minimum mass. Total mass of the clusters is $10^5M_{\odot}$. Summary of naming convention for these simulated clusters is explained hereafter:} 
\label{table:sims}
\hspace*{\fill} $\underbrace {S}_{1}$~$\underbrace {10 ~seg}_{2}$~$\underbrace {03 ~bin}_{3}$~$\underbrace {01 ~m ~100}_{4}$\hspace*{\fill}
\begin{itemize}
\item [1 :] {\bf S} means Stellar evolution is ON and {\bf D} means Stellar evolution is OFF (Pure Dynamically evolution).
\item [2 :] Number before {\bf seg} shows the degree of mass segregation. {\bf 10 seg} means fully segregated and {\bf 00 seg} means non-segregated.
\item [3 :] Number before {\bf bin} shows the initial binary fraction. {\bf 03 bin} means 0.3 binary fraction (30 percent of binaries) and {\bf 00 bin} means no initial binaries.
\item [4 :] The numbers before and after {\bf m} shows the mass range. The first number before {\bf m} can be {\bf 01} which means the low mass cutoff is $0.1 M_{\odot}$ or it can be {\bf 10} which means the low mass cutoff is $1.0 M_{\odot}$ and the second number after {\bf m} can be {\bf 100} which means the maximum mass of the particles is $100 M_{\odot}$ or it can be {\bf 300} which mean the maximum mass is $300 M_{\odot}$.
\end{itemize}
So name {\bf S10seg03bin01m100} stand for a cluster with stellar evolution is ON and it is fully segregated with 0.3 binary fraction and mass range between $0.1 M_{\odot}$ to $100 M_{\odot}$. Also {\bf D00seg00bin10m300} means this cluster evolves just Dynamically without stellar evolution and it is not initially segregated without any initial binaries and mass range between $1.0 M_{\odot}$ to $300 M_{\odot}$.
\end{table*}


\section{Results of the simulations}\label{sec:simresults}
\subsection{Expansion of the cluster}
\begin{figure*}[!]
 \centering 
\includegraphics[width=17.cm]{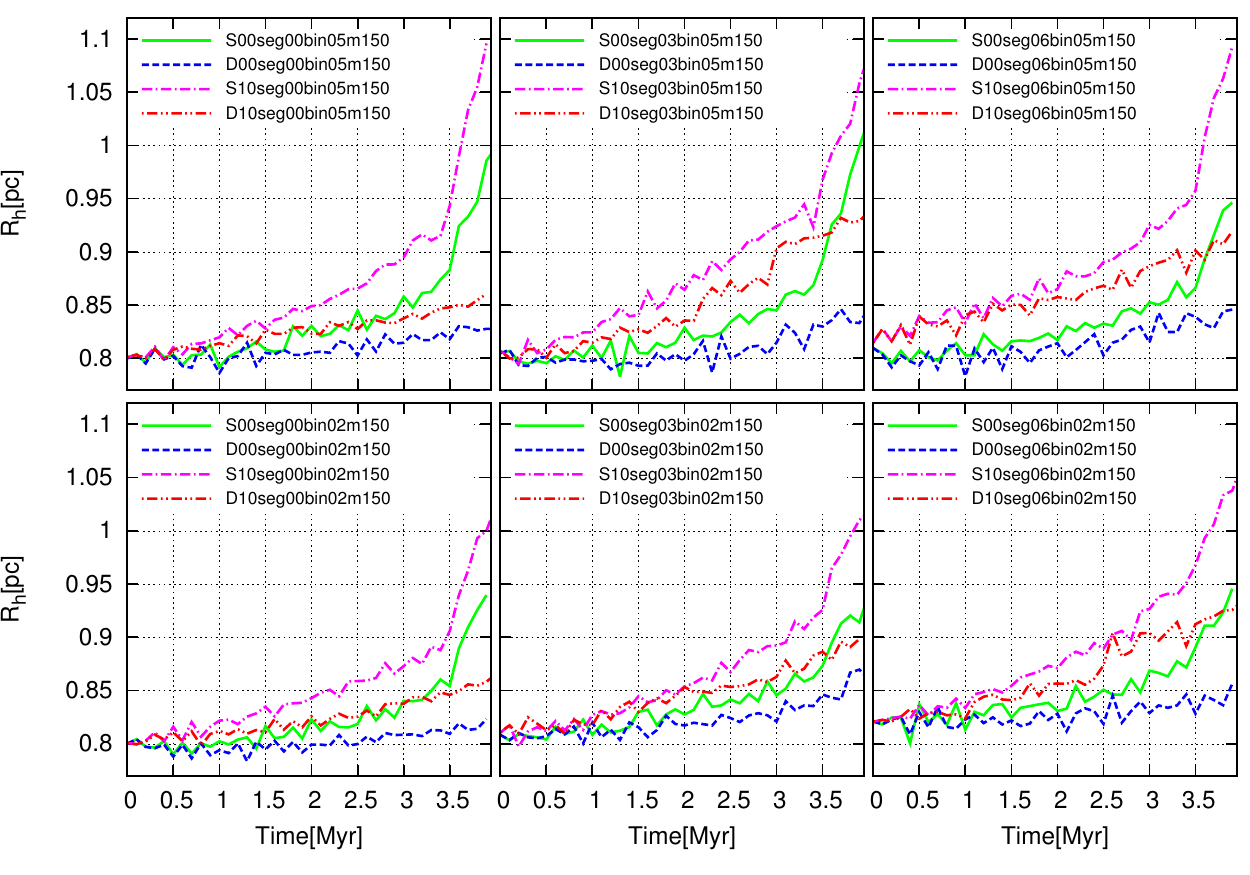}
\caption[]{Half-mass radius evolution of different clusters within 4 Myr. Up: mass range of $0.5 M_{\odot}-150M_{\odot}$. Down: $0.2 M_{\odot}-150M_{\odot}$. Left: No initial binaries, Middle: 30\%  initial binaries, Right: 60\%  initial binaries.  Green and Blue: Non-segregated; Pink and Red: Segregated.
\label{fig:rh150}}
\end{figure*}

Figure \ref{fig:rh150} shows the evolution of half mass radii of 24 simulated clusters in 4 Myr. The upper plots correspond to clusters with a mass distribution in the range of $0.5 M_{\odot}-150M_{\odot}$ and bottom plots correspond to clusters with mass distribution in the range of $0.2 M_{\odot}-150M_{\odot}$. From left to right the plots depict clusters with 0\%, 30\% and 60\% initial binaries.
It can be seen in all plots that segregated clusters expand more than non-segregated ones. Stellar evolution plays also an important role in the expansion of the cluster, especially around 3-3.5 Myr on the evolution of massive stars and their high mass-loss. Clusters which contain more massive stars, present a larger expansion. This can be checked by comparing the half-mass radius evolution of clusters with pure dynamical evolution (D-clusters). At the same time, changing the binary fraction does not affect the expansion of the clusters in a significant way.

\subsection{Escapers and cluster's mass loss}
\begin{figure*}[!]
 \centering 
        \includegraphics[width=9.0cm]{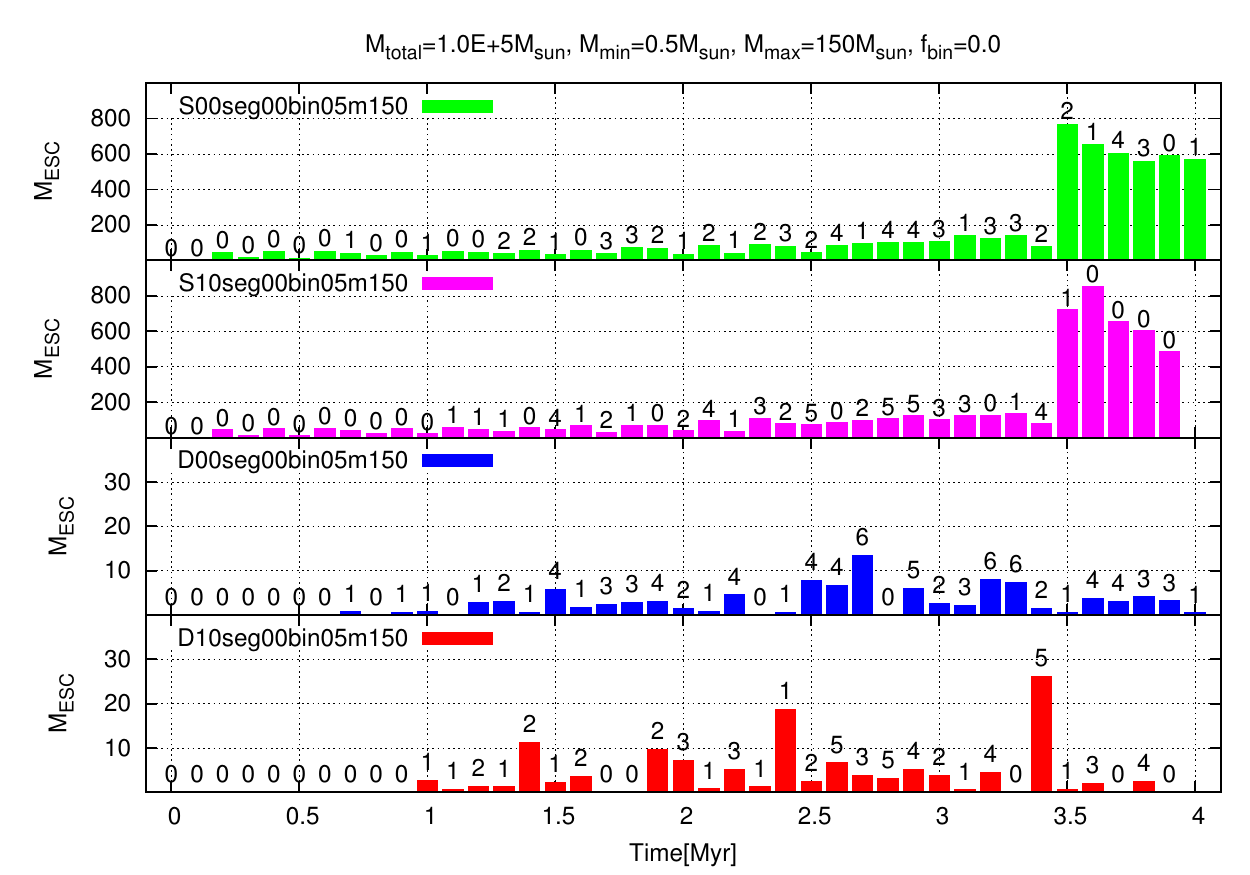}
        \includegraphics[width=9.0cm]{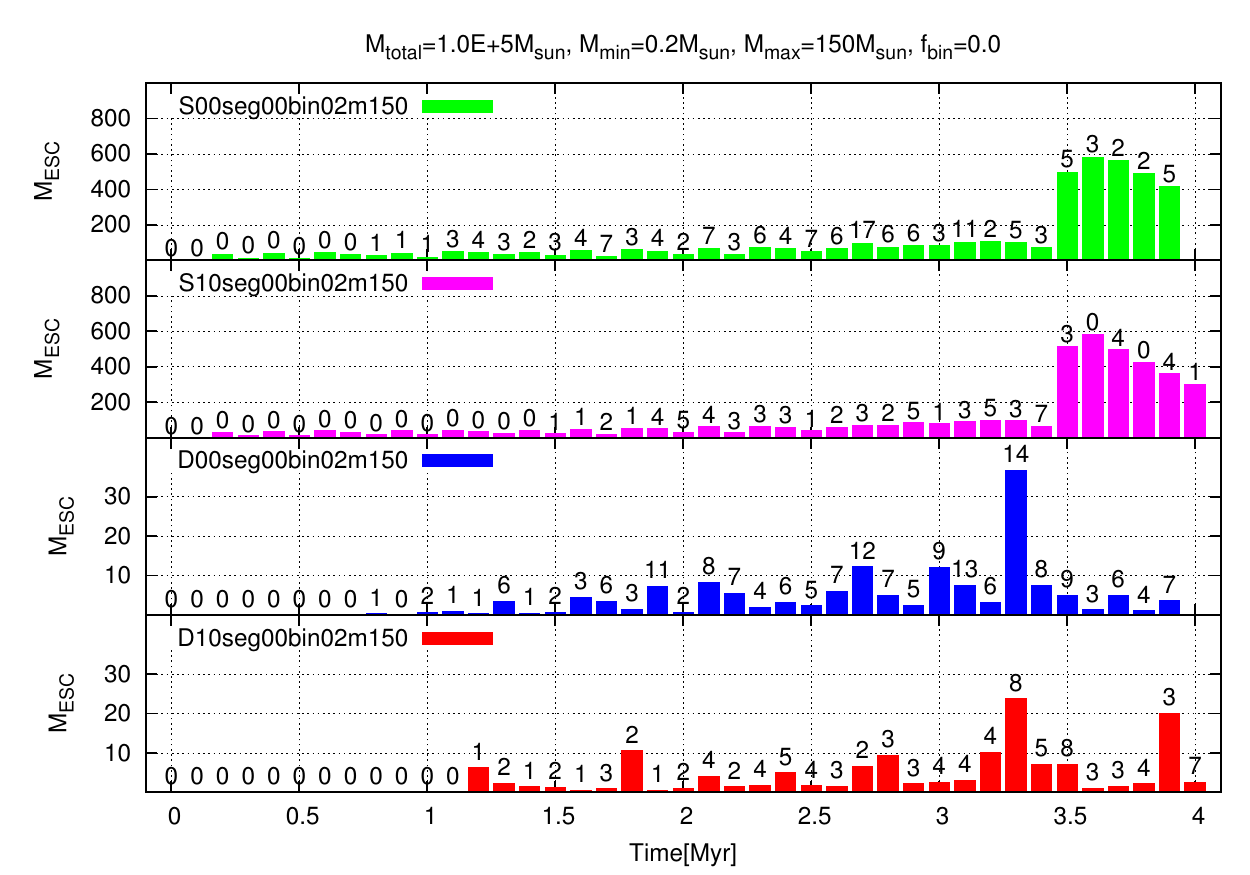}\\
        \includegraphics[width=9.0cm]{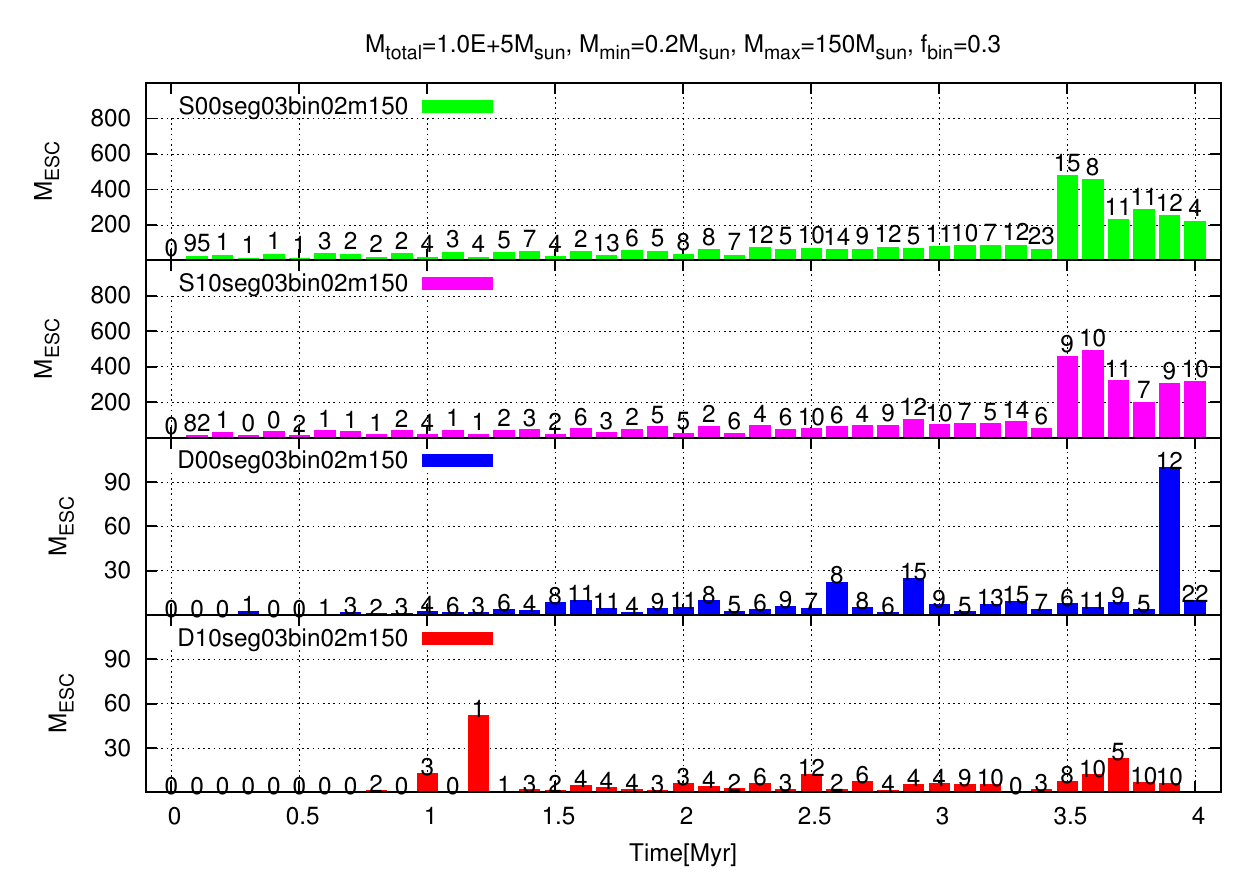}
        \includegraphics[width=9.0cm]{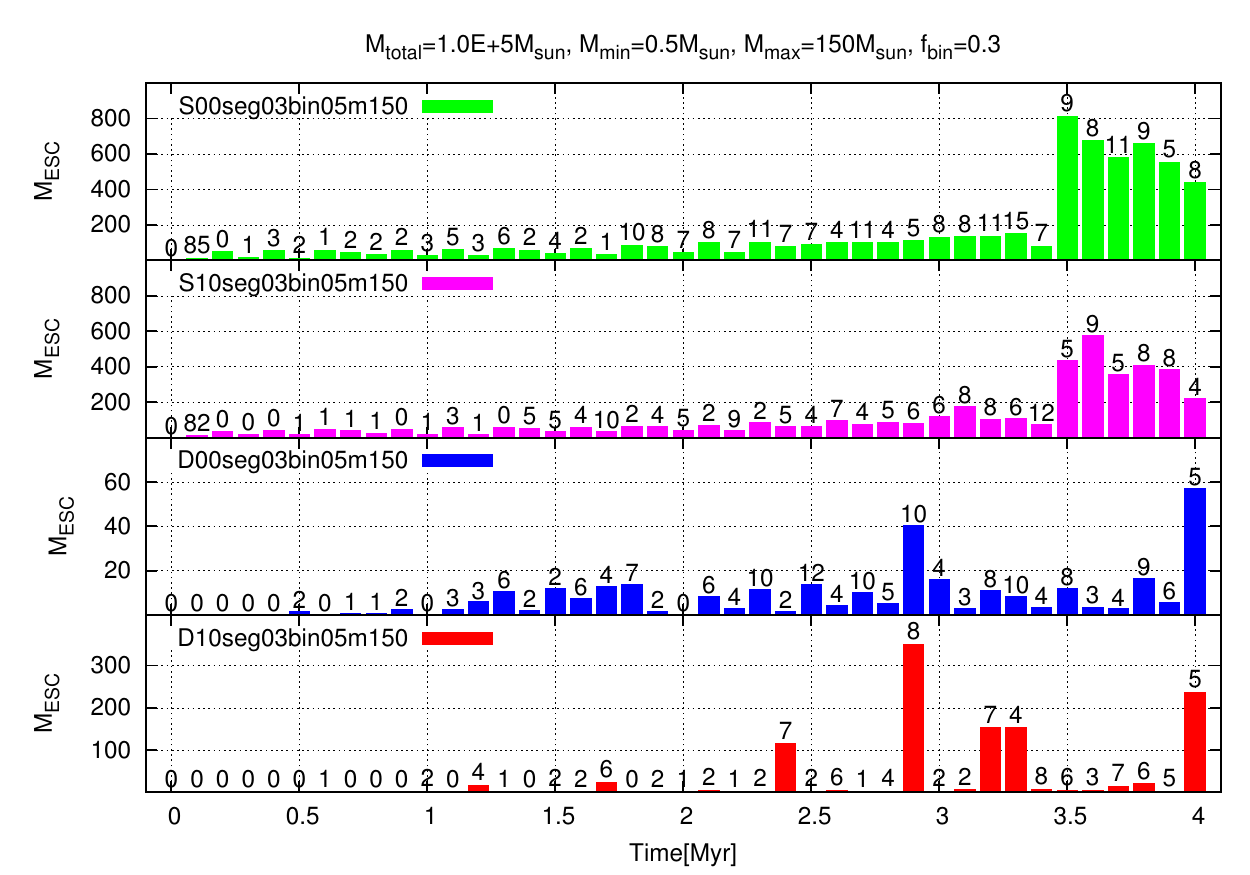}\\
        \includegraphics[width=9.0cm]{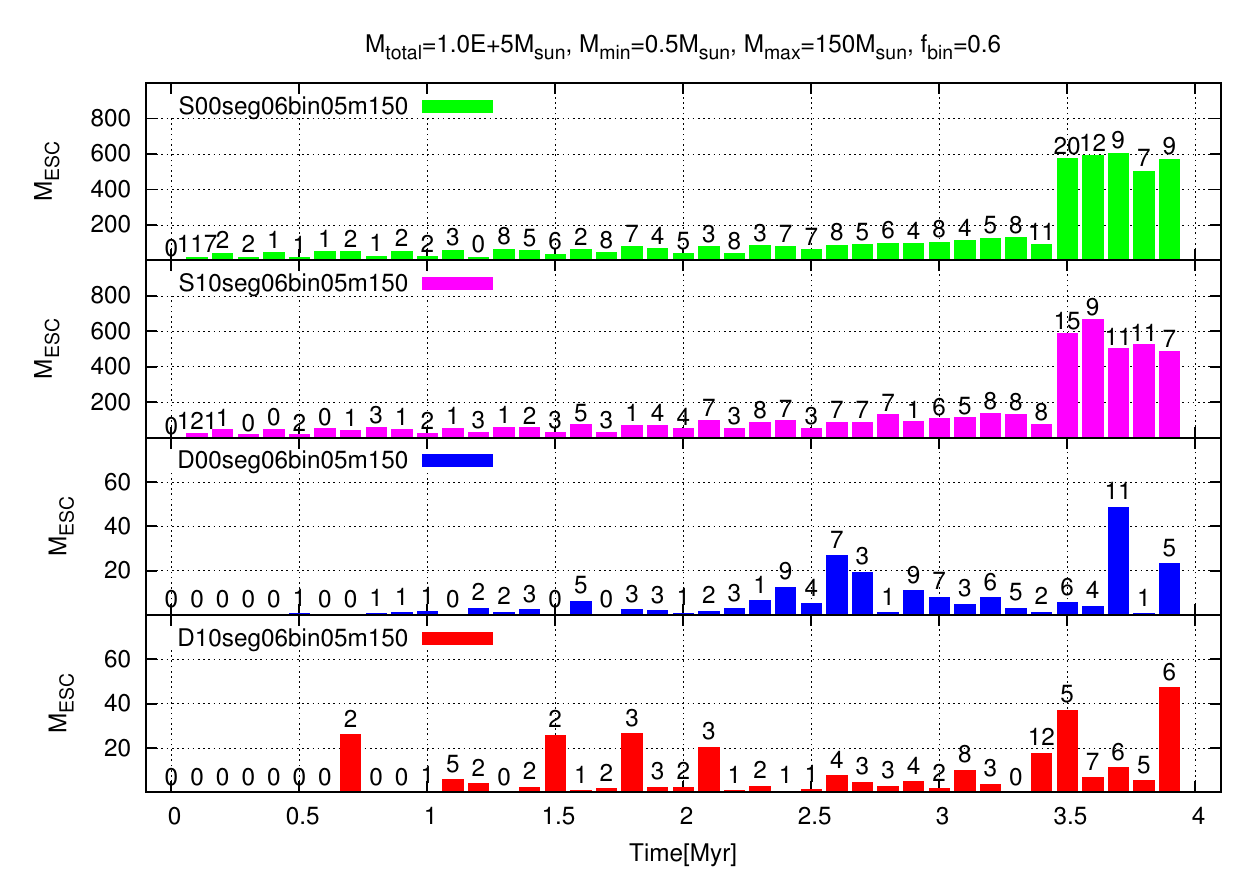}
        \includegraphics[width=9.0cm]{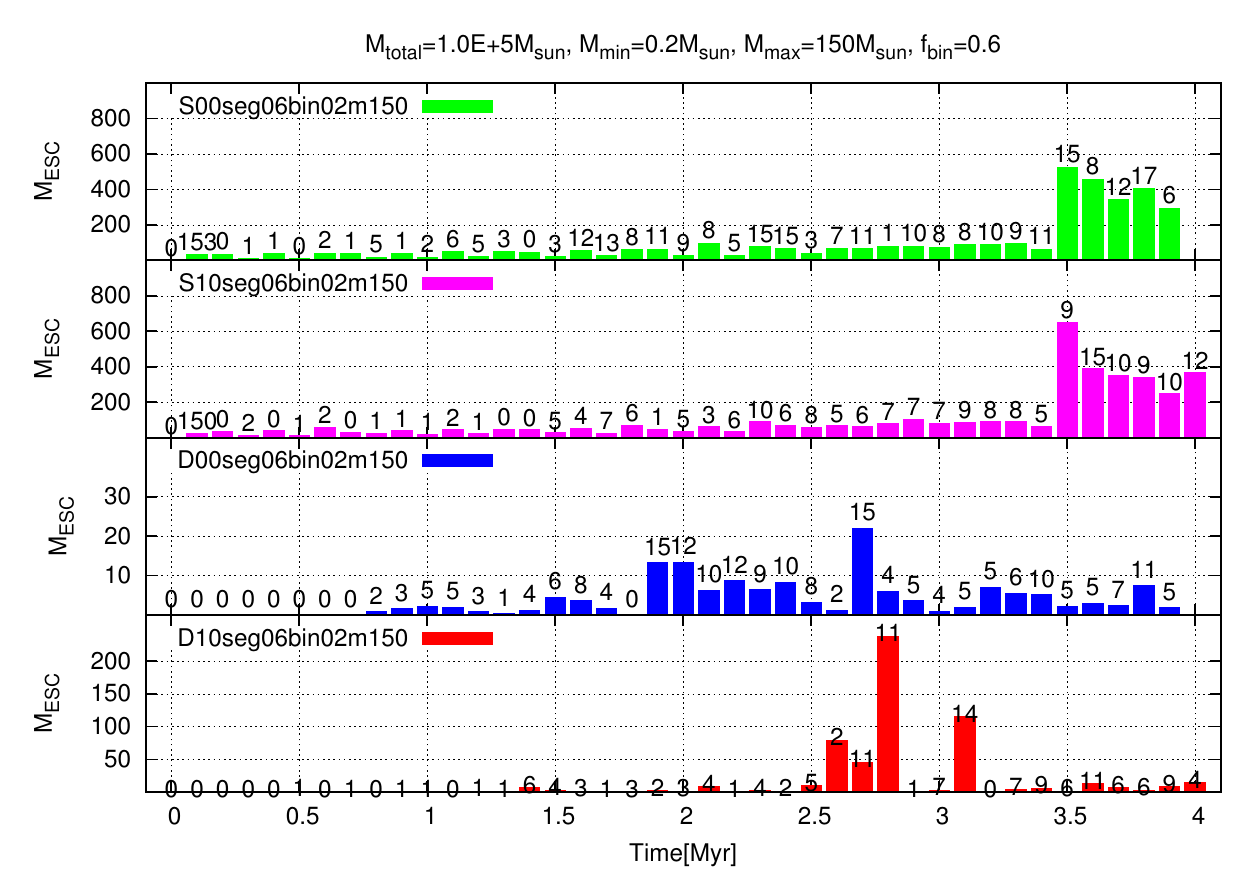}
\caption[]{Total mass loss of the clusters per time-step. Left: mass range of $0.5 M_{\odot}-150M_{\odot}$. Right: mass range of $0.2 M_{\odot}-150M_{\odot}$. Upper, middle and bottom plots represent clusters with 0\%, 30\%  and 60\% initial binaries. The numbers at the top of each bin correspond to the number of escapers.
Green and Blue: non-segregated; Pink and Red: Segregated.
Stellar evolution is ON for Green and Red, and OFF for blue and Red. 
\label{fig:esc150}}
\end{figure*}

Clusters lose mass due to escaping stars and also if stellar evolution is ON, which drives the mass-loss of massive stars. Figure \ref{fig:esc150} shows the total mass loss of each cluster per time-step. Left plots correspond to clusters in the mass range of $0.5 M_{\odot}-150M_{\odot}$ and right plots for $0.2 M_{\odot}-150M_{\odot}$. Upper, middle and bottom plots represent clusters with 0\%, 30\%  and 60\% initial binaries. 

The mass-loss of the clusters with stellar evolution (hereafter call S-clusters) is much larger than the clusters without stellar evolution (here after call D-clusters) especially around time 3.5 Myr which is a time when massive stars ($M > 60 M_{\odot}$) turn out to supernova events. Clusters on the left with $M_{min} = 0.5 M_{\odot}$ contain indeed more massive stars than clusters at the right (with $M_{min} = 0.2 M_{\odot}$), so for these clusters mass loss is more than for clusters with $M_{min} = 0.2 M_{\odot}$.
that for D-clusters the number of escapers increases with the increasing binary fraction.

\subsection{Binary fraction}\label{sec:binary}
\begin{figure*}[!]
 \centering 
\includegraphics[width=17.cm]{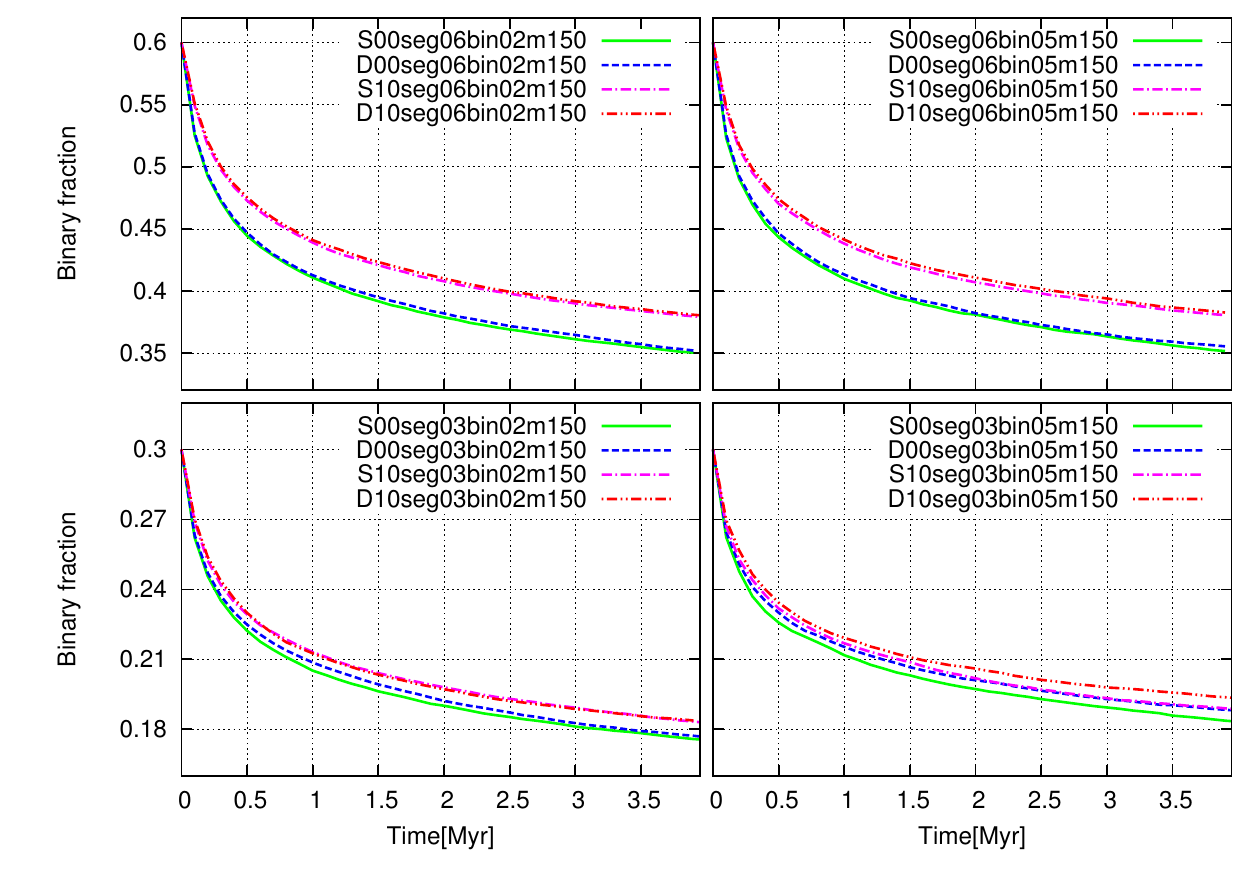}
\caption[]{Fraction of bound binary systems for different clusters in each time-steps. 
Left, Up: clusters with mass range of $1.0 M_{\odot}-100M_{\odot}$ and 30\% Initial binaries.
Right, Up: clusters with mass range of $1.0 M_{\odot}-300M_{\odot}$ and 30\% Initial binaries.
Left, Down: clusters with mass range of $0.5 M_{\odot}-150M_{\odot}$ and 30\% Initial binaries.
Right, Down: clusters with mass range of $0.5 M_{\odot}-150M_{\odot}$ and 60\% Initial binaries.
Green and Blue: non-segregated; Pink and Red: Segregated.
Stellar evolution is ON for Green and Red, and OFF for blue and Red. 
\label{fig:bf}}
\end{figure*}

Figure \ref{fig:bf} shows the fraction of bound binary systems for different clusters versus time.
Segregated clusters lose less binaries than non-segregated clusters.
It seems that the segregated clusters are safer places for binaries to be survive. It can be explain by two main reasons: Location of the binaries and their neighbors.

In segregated clusters binaries are located deeper in the cluster and they interact mostly with the same-mass neighbors so the chance to be disrupted by single massive star and massive binaries decreases in segregated clusters. Also when a binary is disrupted in the segregated cluster, If it is going to be ejected/evaporated from cluster, It has to pass from a outer layer of the cluster which is contaminated by single stars. This star still has a chance to interact with single stars and remain in the cluster which decreases the evaporation probability. This is not a case for Non-segregated clusters.

For clusters with 30\% binaries it can be seen that if they contain more low mass stars and less massive binaries (the case of clusters with $M_{min} = 0.2 M_{\odot}$), they lose more binaries than clusters with $M_{min} = 0.5 M_{\odot}$.

\subsection{Periods and eccentricities}
\begin{figure*}
 \centering 
        \includegraphics[width=9.cm]{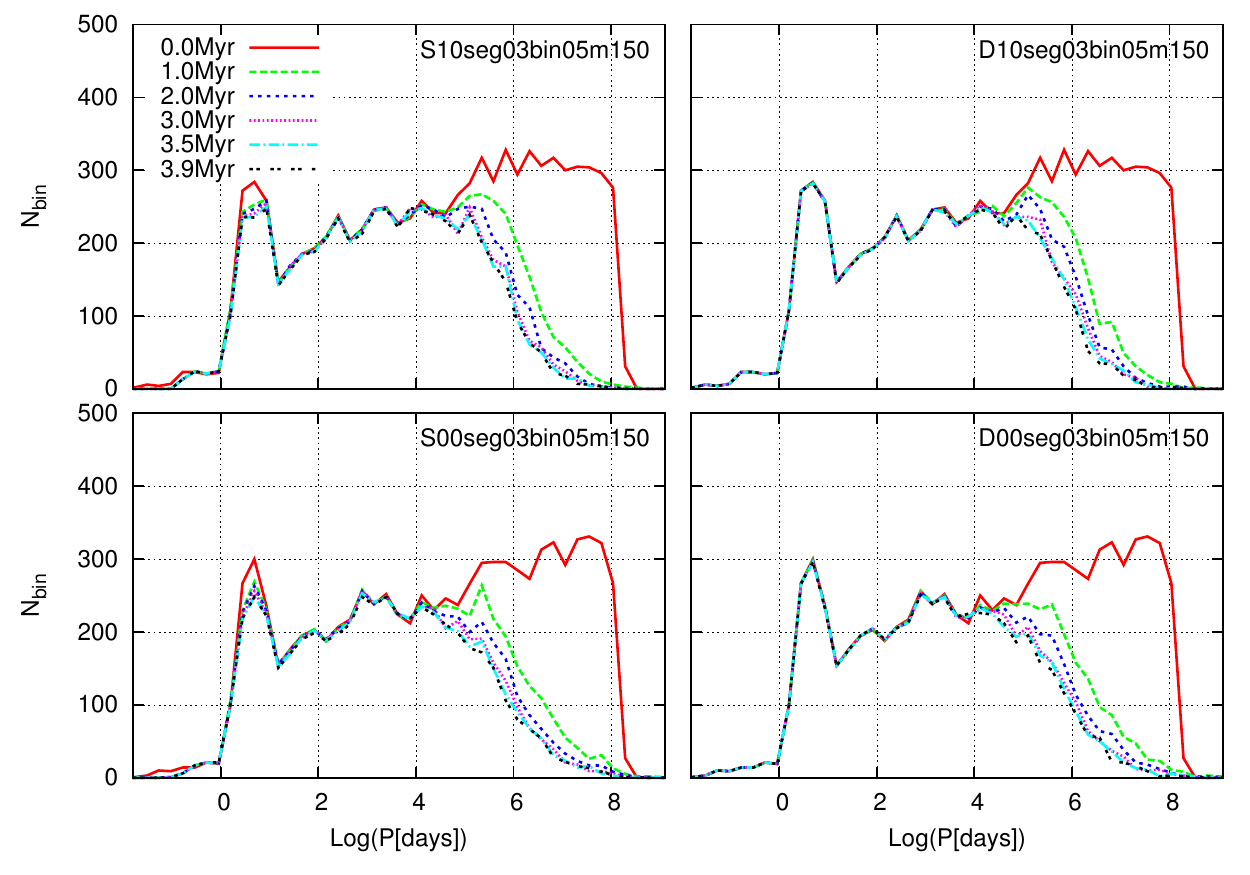}
        \includegraphics[width=9.cm]{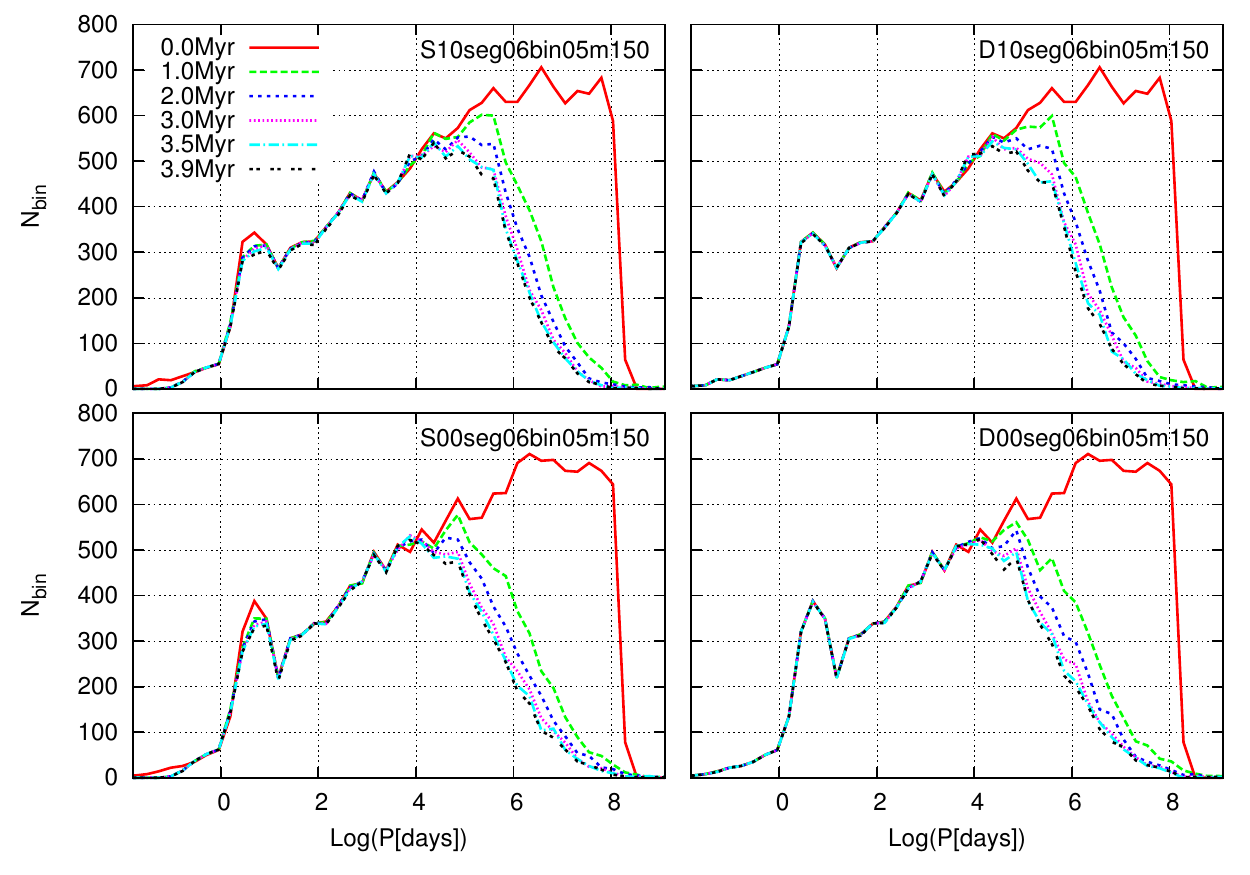}
        \includegraphics[width=9.cm]{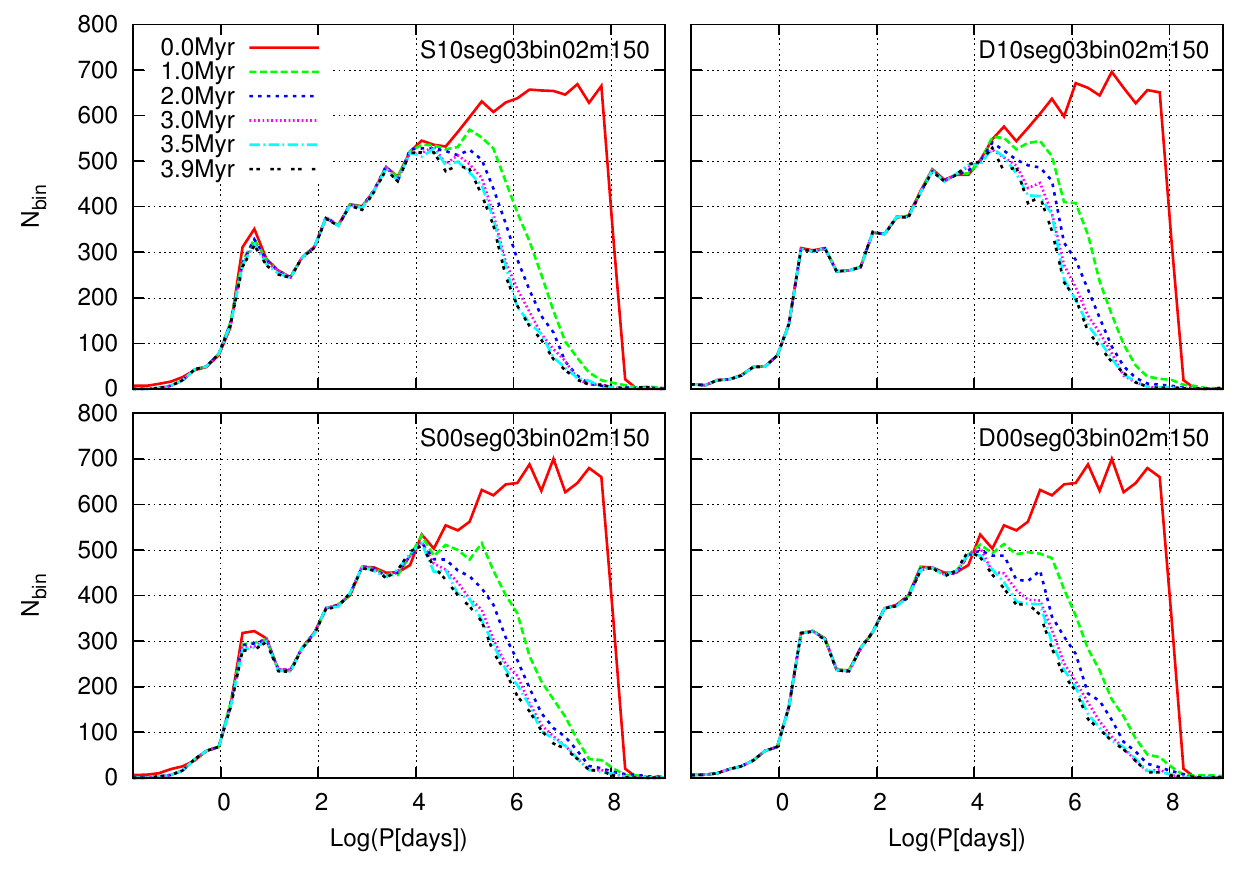}
        \includegraphics[width=9.cm]{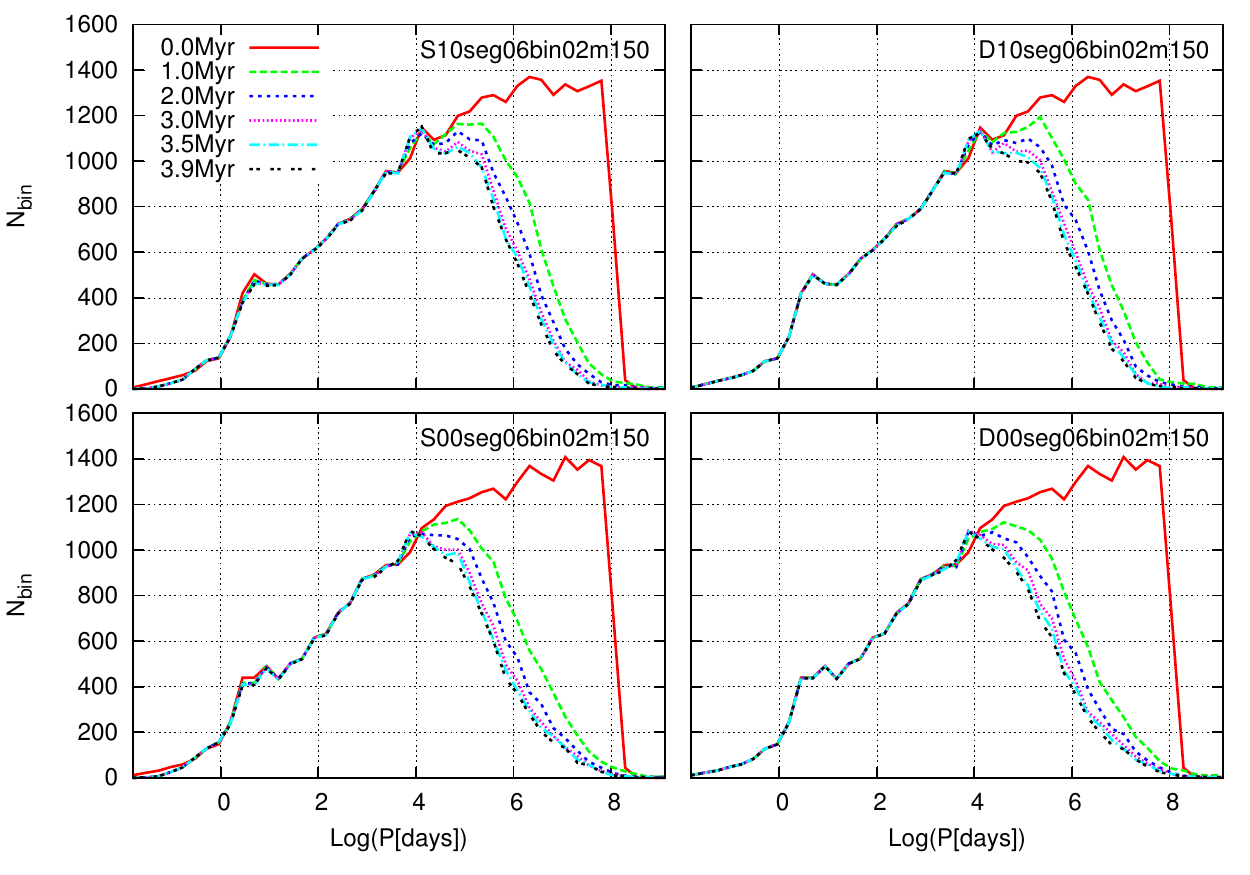}
\caption[]{Histogram of the Log(period [days]) of bound binary systems for different clusters in 6 time-steps (0.0, 1.0, 2.0, 3.0, 3.5 and 3.9 Myr). 
Left, Up: 4 clusters with mass range of $0.5 M_{\odot}-150M_{\odot}$ and 30\% Initial binaries.
Right, Up: 4 clusters with mass range of $0.5 M_{\odot}-150M_{\odot}$ and 60\% Initial binaries.
Left, Down: 4 clusters with mass range of $0.2 M_{\odot}-150M_{\odot}$ and 30\% Initial binaries.
Right, Down: 4 clusters with mass range of $0.2 M_{\odot}-150M_{\odot}$ and 60\% Initial binaries. 
\label{fig:binlogp}}
\end{figure*}

Figure \ref{fig:binlogp} shows the histogram of periods in units of days (logarithmic) in six time-steps for 16 clusters which contain 30\% and 60\% initial binaries. Different colors represent different times. For example red plot is at the time of 0.0 Myr which is the initial distribution of periods that we have chosen according to observations reported in literature (see Section \ref{sec:initial}). It is a bimodal distribution, for low mass binaries it is Kroupa distribution and for massive O, B binaries it is Sana\&Evans 2011 which exhibits in its a first part a peak around 10 days.

Evolution of the first part is according to stellar evolution, that is why it is not visible in D-clusters. Evolution of second part is according to the dynamics of the cluster, that is wht we see this for both S-clusters and D-clusters.

Almost half of the low-mass binaries dissolve within 1 Myr. 

\begin{figure*}
 \centering 
        \includegraphics[width=9.cm]{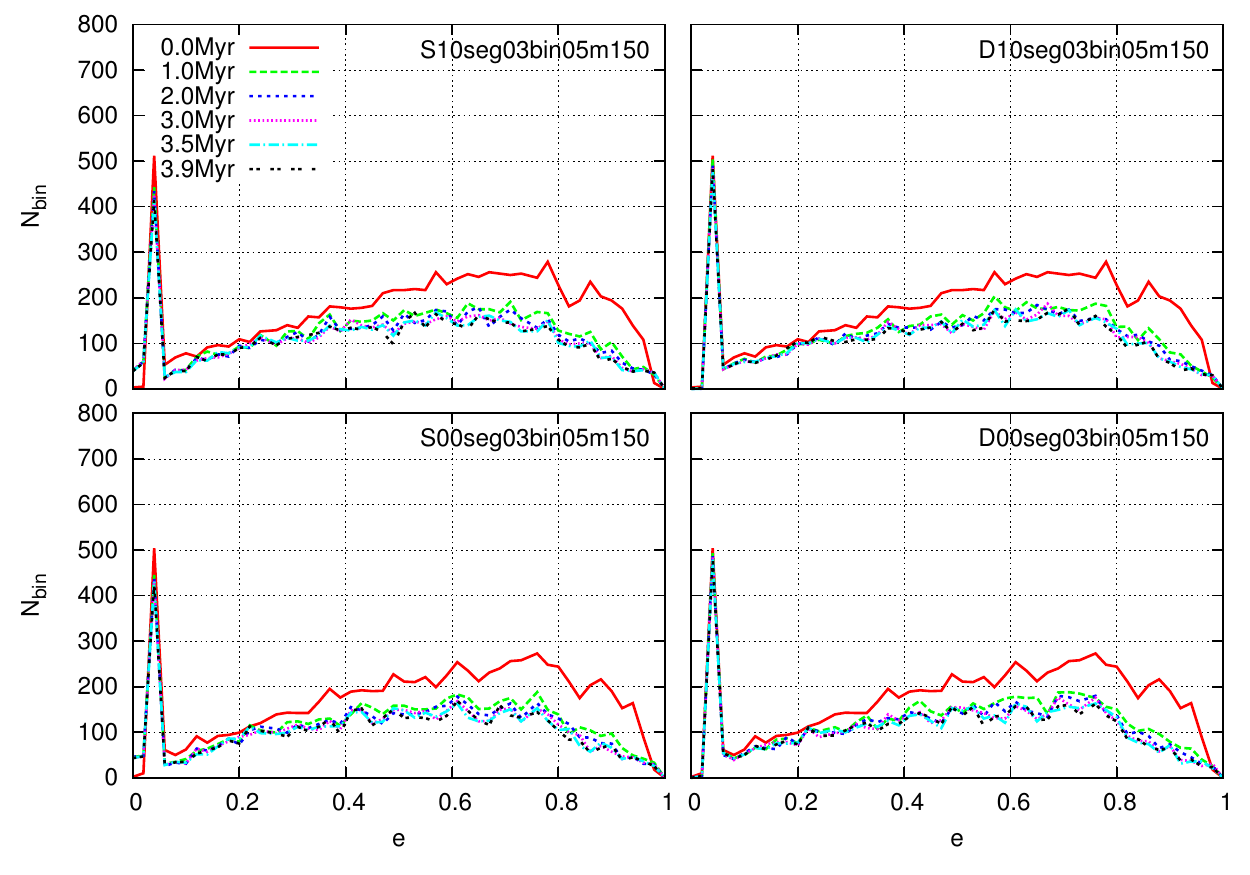}
        \includegraphics[width=9.cm]{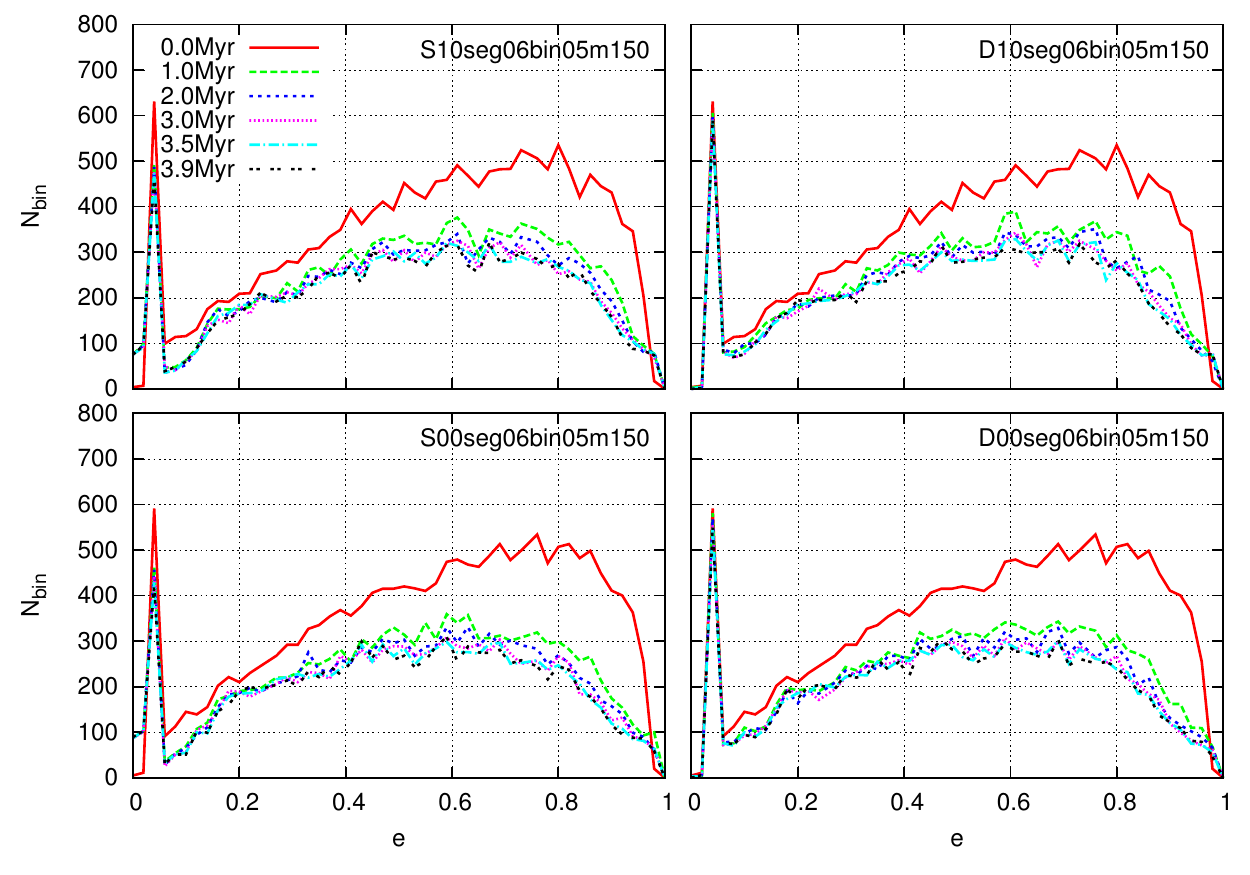}
        \includegraphics[width=9.cm]{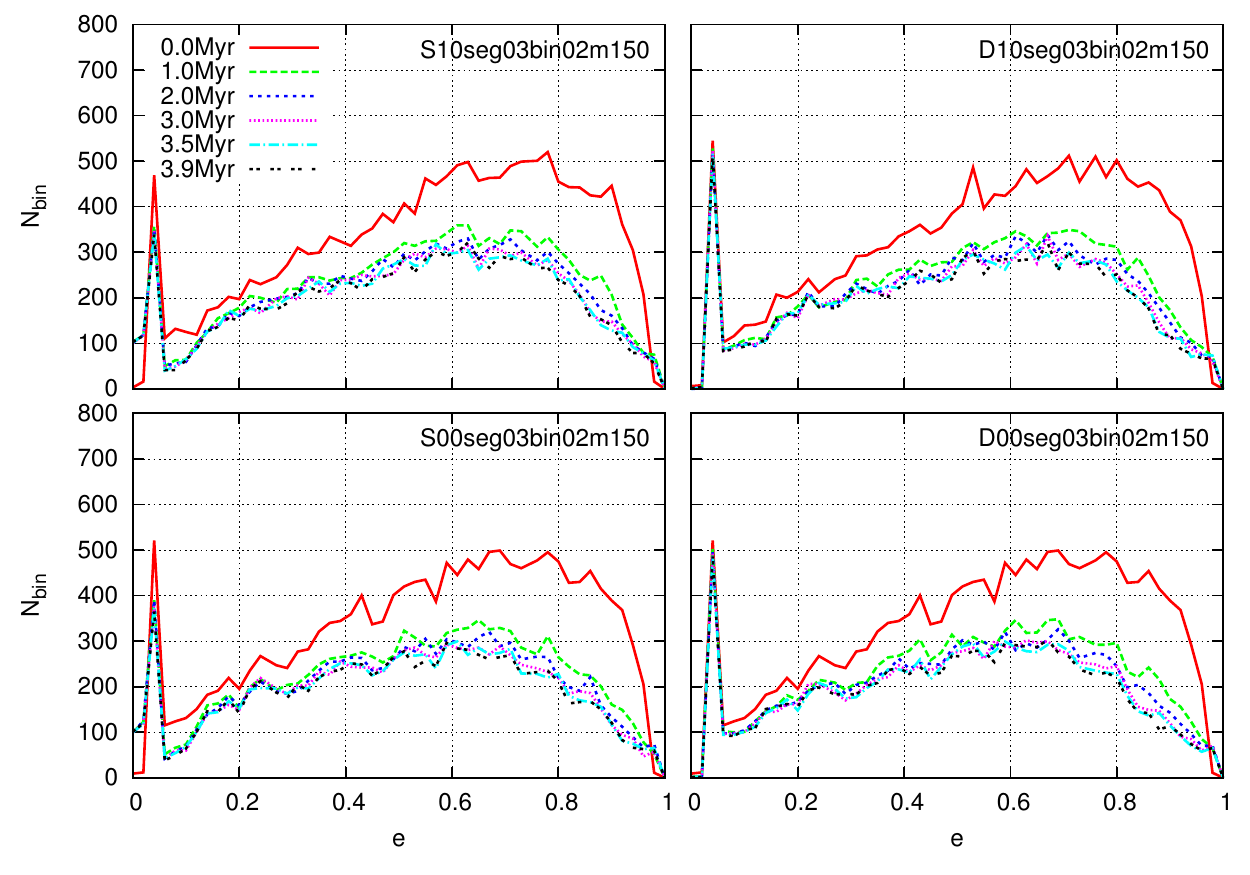}
        \includegraphics[width=9.cm]{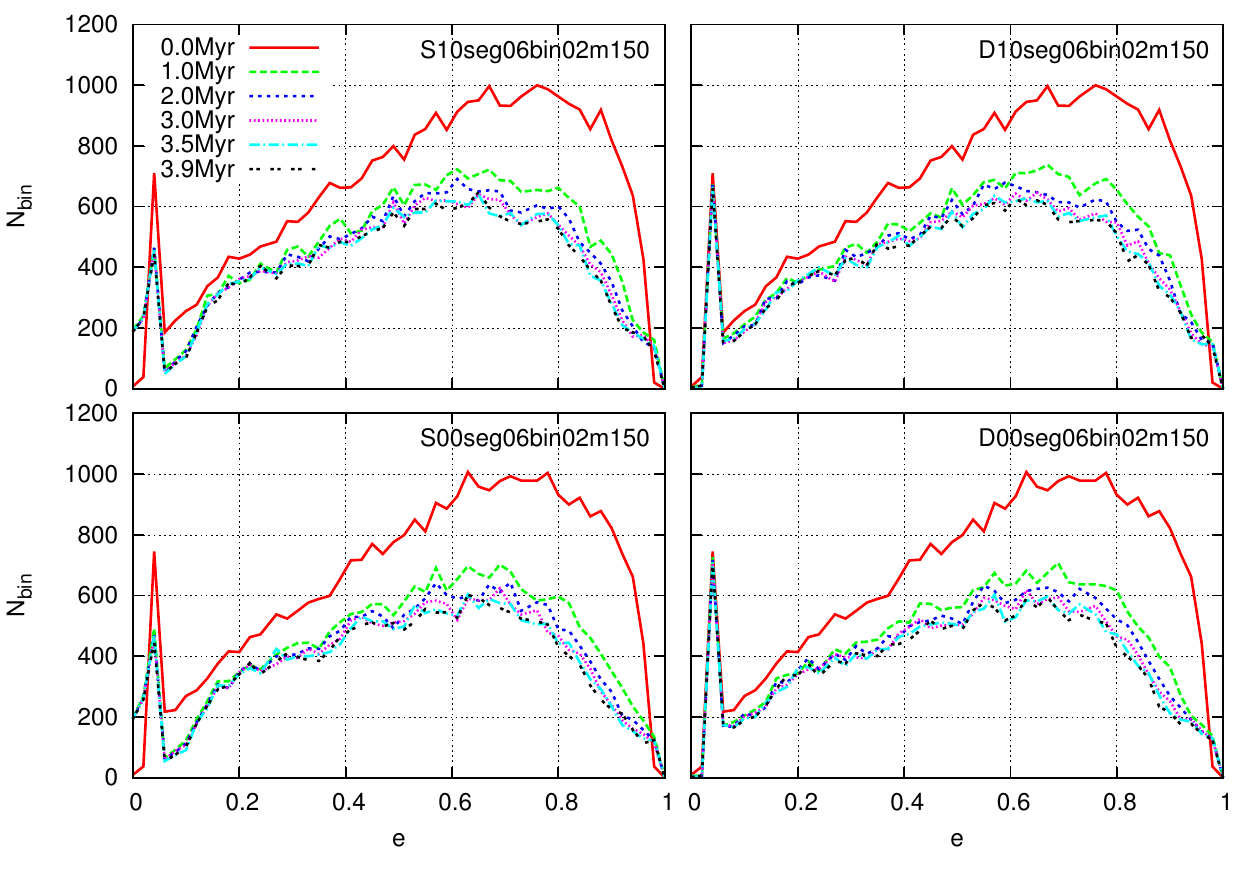}
\caption[]{Histogram of the eccentricity of bound binary systems for different clusters in 6 time-steps (0.0, 1.0, 2.0, 3.0, 3.5 and 3.9 Myr). 
Left, Up: 4 clusters with mass range of $0.5 M_{\odot}-150M_{\odot}$ and 30\% Initial binaries.
Right, Up: 4 clusters with mass range of $0.5 M_{\odot}-150M_{\odot}$ and 60\% Initial binaries.
Left, Down: 4 clusters with mass range of $0.2 M_{\odot}-150M_{\odot}$ and 30\% Initial binaries.
Right, Down: 4 clusters with mass range of $0.2 M_{\odot}-150M_{\odot}$ and 60\% Initial binaries. 
\label{fig:bine}}
\end{figure*}

Figure \ref{fig:bine} shows the evolution of eccentricity distributions in 6 time-steps for 16 clusters which contain 30\% and 60\% initial binaries. Initial distribution (T = 0.0 Myr) is the red plot. Like the period distribution, eccentricities also have a bimodal distributions, for low-mass and massive binaries (with a peak close to e=0.0). Stellar evolution, affects the evolution of the first peak (for massive binaries) that is why for D-clusters the peak does not evolve. For low-mass binaries, during the evolution (in different time-steps) the distribution keeps the memory of the initial distribution for different eccentricities.

For period distribution, after 2-3 Myr the new distribution could keep the memory of initial distribution of massive binaries not for low-mass binaries. But for eccentricity distribution, cluster keeps the memory of initial distribution of eccentricities.

\section{Comparison with observations}\label{sec:compareobs}
For R136, the main observational data result from imaging in different filters. From these data, one can estimate the mass of stars and compare the density profile with simulations. In such an approach errors can dramatically increase from converting magnitudes to mass as the theoretical evolutionary models may not provide enough information for very massive and Wolf-Rayet stars specially. On the other hand, we probably cannot detect many low mass stars during the photometry. Moreover, if the detected object is a binary then estimated mass is biased.

At this point of our study we prefer to produce the imaging data from the simulations with the spatial resolution of HST/WFPC2 data from R136. HST/WFPC2 observations of R136 were carried out on 1994-09-25 (PI: Westphal), from which
we use a combination of shallow, intermediate and long exposures ranging from 3--5\,s
to 80--120\,s with the planetary camera (PC) and the F814W filter. 
So we wrote a code which reads the information of stars from the NBODY6 simulations as an input and creates the synthetic scenes that mimic HST/WFPC2 resolution in different HST filters. During this simulation we also need proper stellar evolution models and atmospheric models for calculating Bolometric Correction (BC) of different HST/WFPC2 filters. We created sets of BC tables at the age of 2 Myr using Geneva stellar evolution models \footnote{http://webast.ast.obs-mip.fr/equipe/stellar/} \citep{geneva}. For calculating BCs we used SEDs from TLUSTY atmosphere models for O and B stars \footnote{Model atmospheres and source codes are
available at http://nova.astro.umd.edu}(\citep{tlusty95},\citet{tlustyO},\citet{tlustyB}) and KURUCZ\footnote{ATLAS9 Kurucz ODFNEW /NOVER models}(\citet{kurucz97}) for the rest of the stellar types with a half-solar metallicity appropriate for LMC stars. 
TLUSTY provides grids of non-LTE, metal line– blanketed, plane-parallel, hydrostatic model atmospheres which is well suited for the very massive stars specially in visible and near-IR.

At a given time (2 Myr) it is possible to calculate the flux (in different HST filters) for each star in the simulation using computed BC tables. We simulated a 800x800 pixels scene corresponding to a field of 32.5"x32.5" on the detector where a star with a given flux falls on the detector with a Gaussian distribution as the PSF profile.

\begin{figure*}
\centering 
        \includegraphics[width=17.cm]{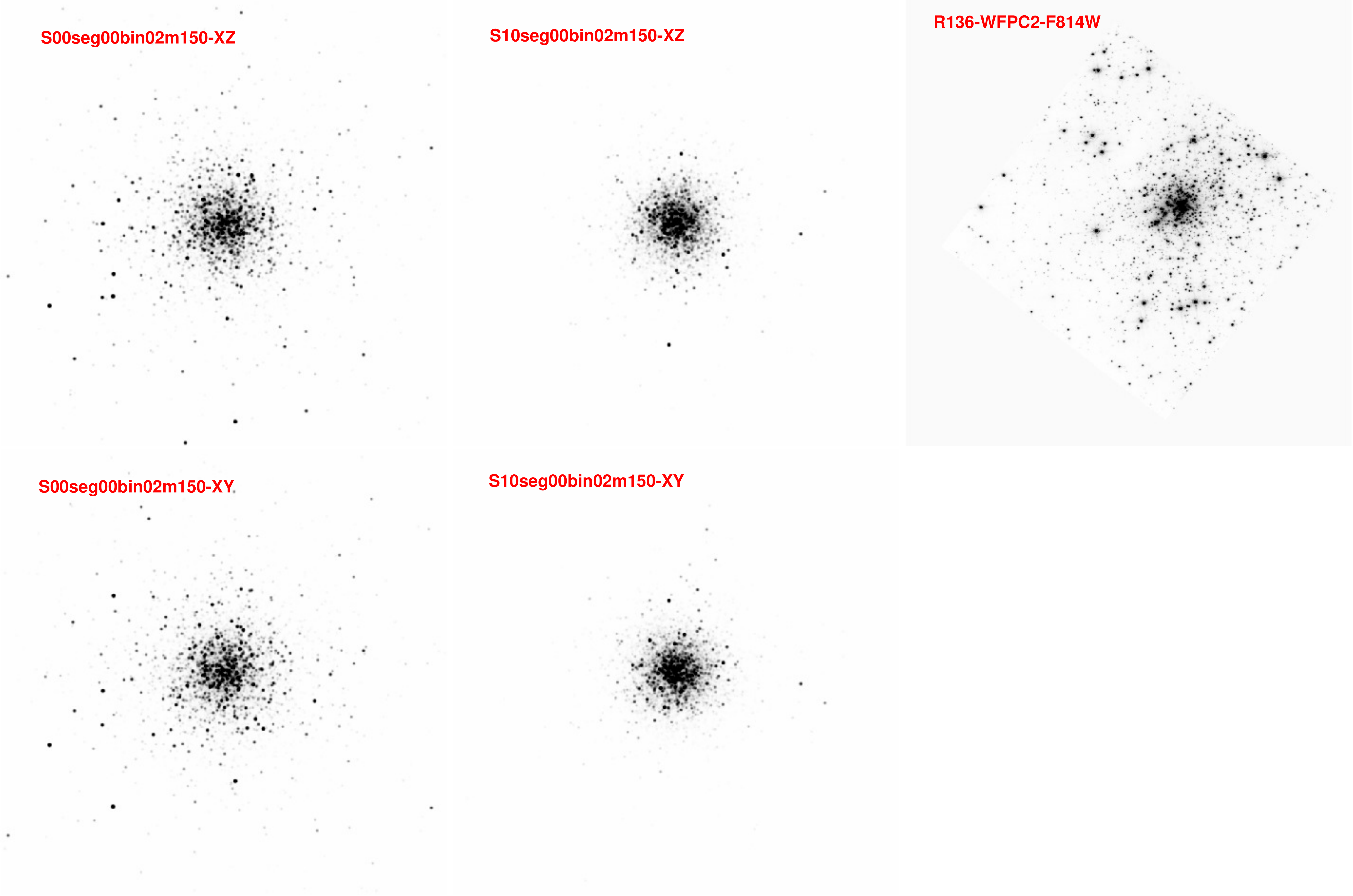}
\caption[]{Synthetic scenes from simulation of S-clusters with the mass range of $(0.2-150) \rm{M}_{\odot}$ and 0\% initial binaries at time 2 Myr. Not segregated: Left images; Segregated clusters: Middle images;  Top is in XZ plane and bottom is in XY plane. R136 image taken by HST/WFPC2-PC in F814W filter is in the Right top. 
\label{fig:scene_r136}}
\end{figure*}

Figure \ref{fig:scenesxz05m150}-\ref{fig:scenesxy05m150} show synthetic scenes of 12 simulated clusters at time 2 Myr both in XY and XZ plane. One can compare the synthetic images with real HST/WFPC2 data on R136 shown in Figure \ref{fig:scene_r136}. Both images are in F814W filters.

However the question is at what degree a simulated cluster with thousands of stars within a given volume projected once on the sky to  reproduce HST image of R136. What is the best criterion to select the closest simulated cluster to R136?

One useful way is to compare the Surface Brightness Profile (SBP) of R136 to those of synthetic scenes (Section \ref{sec:sbp}). It is also possible to compare the Half-Light radius ($\rm{R}_{hl}$) of R136 and synthetic scenes (Section \ref{sec:rhl}). In Section \ref{sec:mfslopes} we compared the Mass Function (MF) slopes of R136 with the MF slopes from simulations.  The MF is not directly derived from simulation. The mass of stars in the FoV is estimated by the photometry on the synthetic scenes and we used the BC-tables for finding the mass of each detected star in a given field.
In Section \ref{sec:neighbor} we introduce a new definition for double checking. In this section we calculate a neighbor radius ($R_{neighbor}$) of each star in each cluster which is a radius containing for example, 100 neighbor stars. In a crowded regions (in the core) this radius is very short for each star while in outer regions it can be larger.

\subsection{SBP of R136}\label{sec:sbp}
\begin{figure*}
 \centering 
        \includegraphics[width=10.cm]{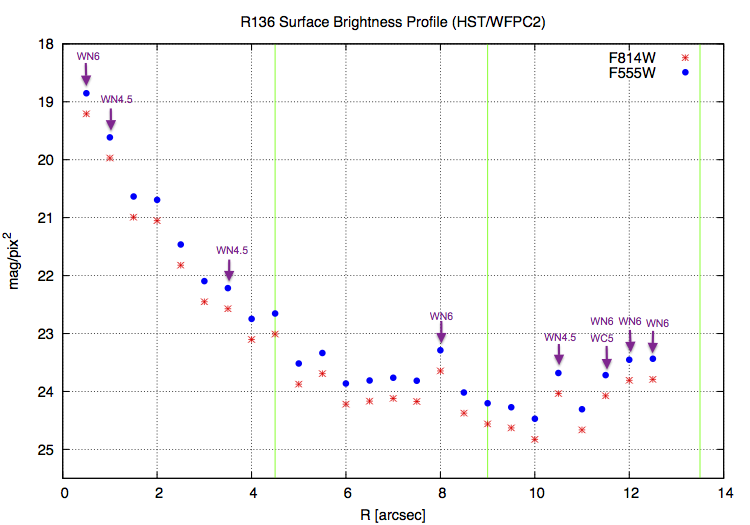}
\caption[]{SBP of R136 in two filters F814W and F555W. 9 WR stars are shown in the plot. 
\label{fig:wr}}
\end{figure*}

Figure \ref{fig:wr} shows the SBP of R136. It can be seen that at some radial distances from the core of the cluster the SBP suddenly increases with a significant deviation from the general trend. We checked for the distribution of the spectral type of R136 stars in the FoV of HST/WFPC2-PC imaging data and we found 9 WR stars . Figure \ref{fig:wr} shows the radii at which a given WR is detected.

\begin{figure*}
 \centering 
        \includegraphics[width=8.cm]{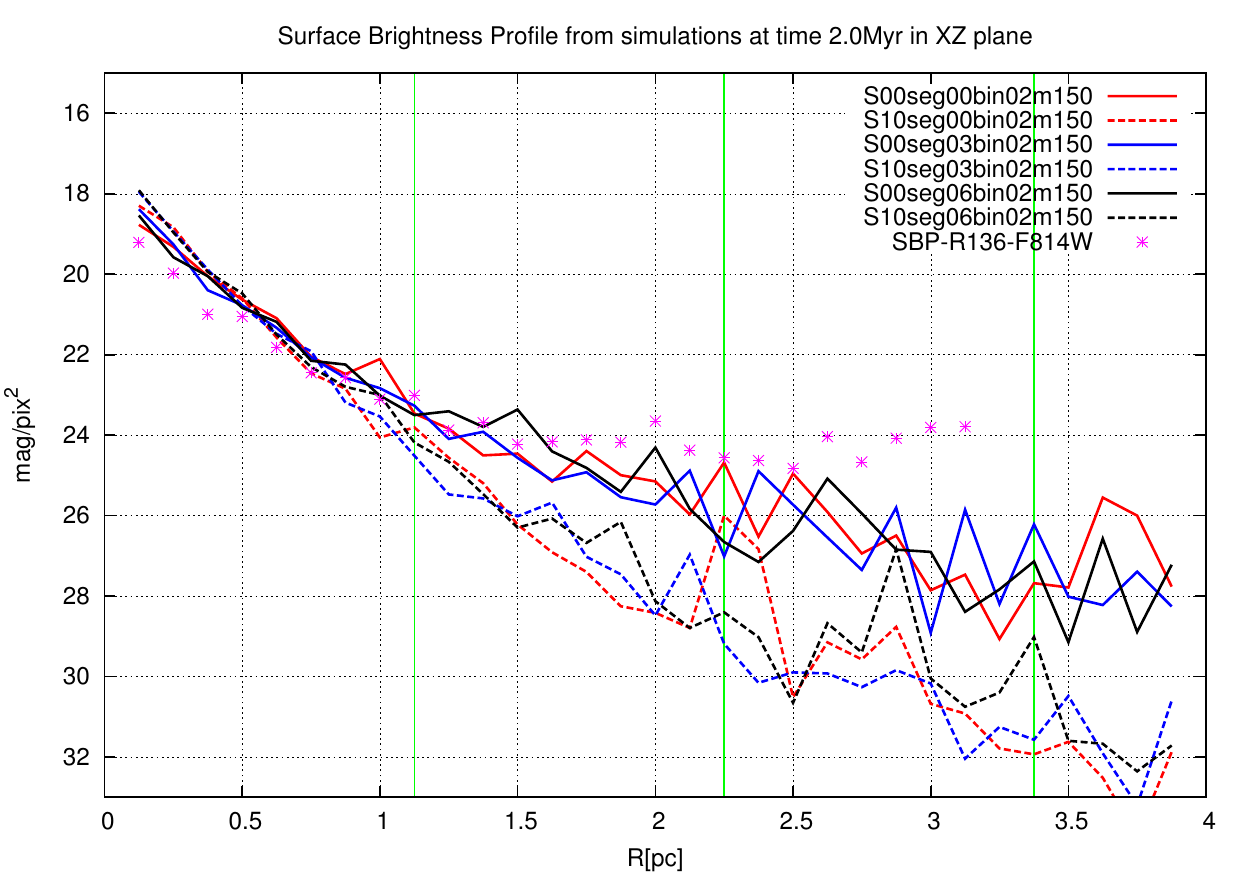}
        \includegraphics[width=8.cm]{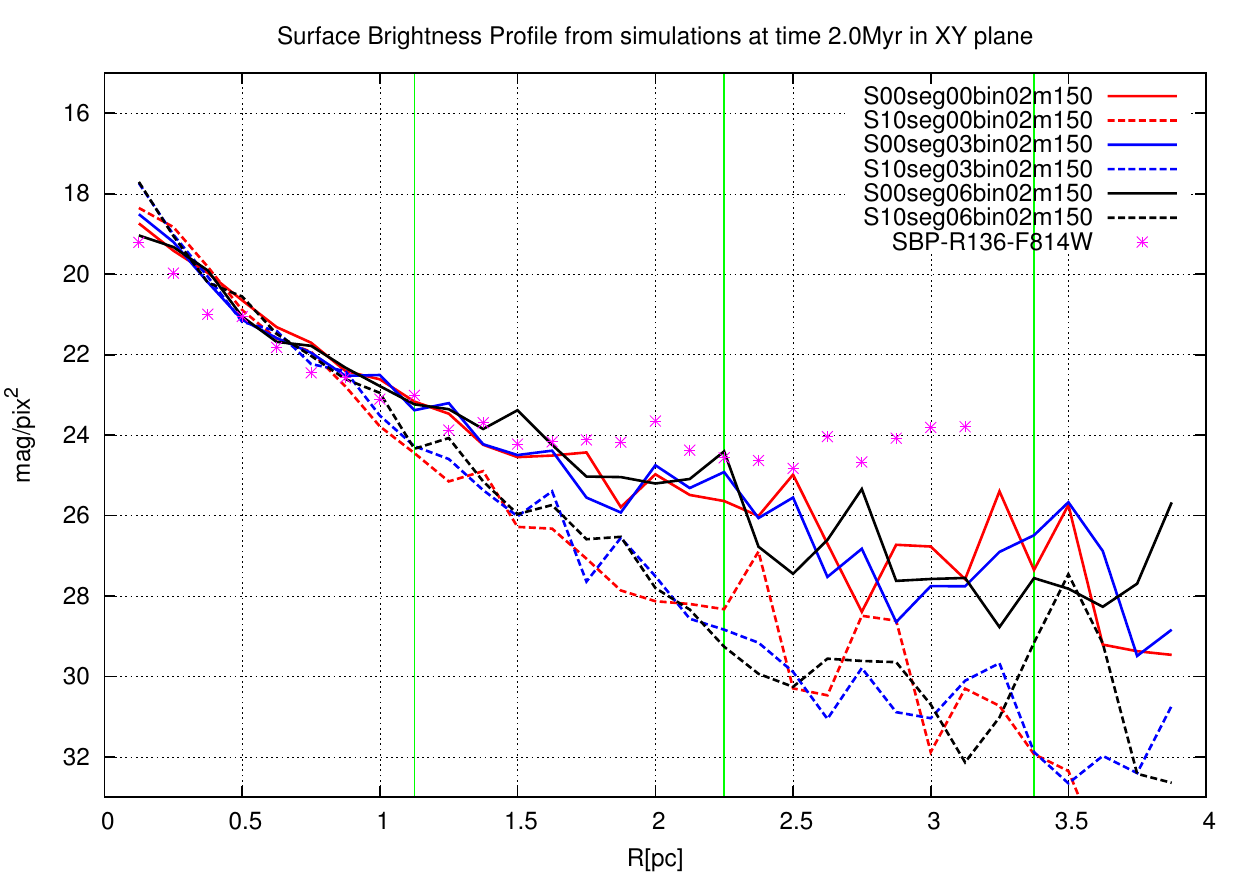}\\
        \includegraphics[width=8.cm]{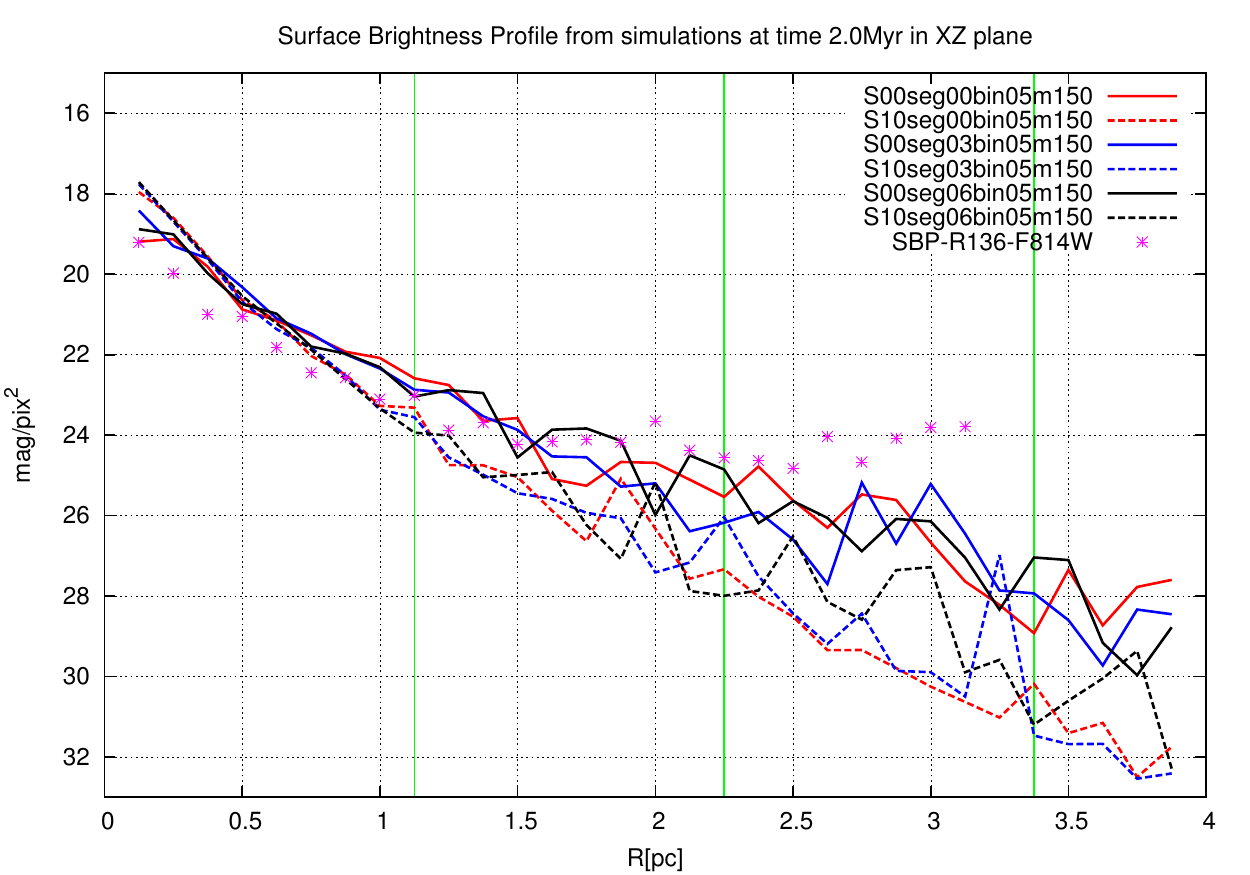}
        \includegraphics[width=8.cm]{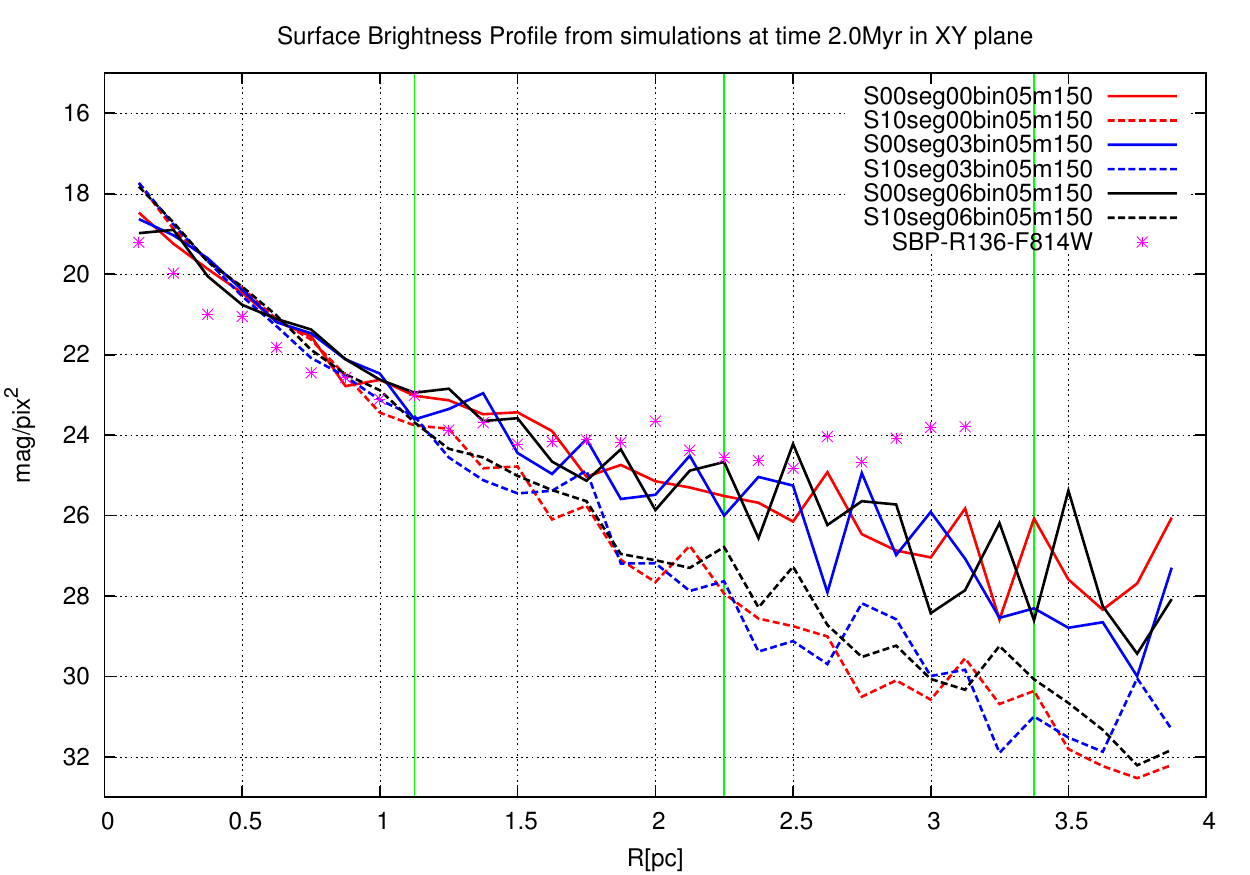}
\caption[]{SBP from simulations at time 2 Myr for different sets of clusters in a mass range of $(0.2-150) \rm{M}_{\odot}$ (upper plots) and $(0.5-150) \rm{M}_{\odot} $ (bottom plots). Right is the synthetic scenes created in XY plan and Left belongs to XZ plan. Solid lines belong to initial non-segregated clusters and dashed lines represent initial segregated clusters.

Pink stars shows the SBP of R136 from WFPC2-PC data. Reddening is corrected for the HST data.
\label{fig:sbpsim}}
\end{figure*}

For synthetic scenes we have calculated the SBPs in the same regions as R136. Figure \ref{fig:sbpsim} shows the SBPs of the simulated R136 versus its simulated twin (pink stars). 

To determine the closest SBP value of the simulated to the observed R136 we calculated the $\chi^2$ for each cluster in three regions in addition to the whole cluster. Table \ref{table:k2sbp-xz} (simulated scenes in XZ plane) and Table \ref{table:k2sbp-xy}  (simulated scenes in XY plane) depict the $\chi^2$ values. Clusters with the smallest value have an SBP more closer to that of R136. Thus the non-segregated clusters match better R136 in all regions.

One can consider the cumulative SBP of R136 and compare it with cumulative SBP of R136. As this cumulative value smooths the SBP, it hides the details of the profile but we calculate this function for both R136 HST data and simulations. The results are given as an example in Appendix \ref{sec:cumSBP}.

\begin{table*}
\centering 
\begin{tabular}{|c c c c c c c |} 
\hline
 & & &$0.2 - 150 M_{\odot}$ & & & \\
\hline
 & &Non segregated & & & Segregated & \\
\hline
Binary:  &  0\% & 30\% & 60\% & 0\% & 30\% & 60\%\\ 
\hline
reg1:  &   0.165324  &   0.105162  &   0.113661  &   0.243965  &   0.338743  &   0.275119   \\
reg2:   &    0.310922  &   0.592480   &  0.420130  &   3.60058  &    3.51407    &  3.23245     \\
reg3:  &  2.00193  &    2.23264   &   2.25797   &   8.67890    &  11.7996   &   8.54216    \\
total:    &  2.46867    &  2.68223    &  2.60257  &    12.4118    &  14.6848   &   11.3888   \\
\hline
 & & &$0.5 - 150 M_{\odot}$ & & & \\
\hline
 & &Non segregated & & & Segregated & \\
\hline
Binary:  &  0\% & 30\% & 60\% & 0\% & 30\% & 60\%\\ 
\hline
reg1:  &  0.234170  &  0.278665   &  0.199646   &  0.315313   &  0.327533   &  0.378822\\
reg2:  & 0.282935   &  0.481584   &  0.310920   &  1.56187  &    1.53784   &   1.77578\\
reg3:  &  1.36780    &  1.53290   &   1.33918   &   8.44827   &   7.46168     & 4.86113\\
total:    &   1.83876    &  2.18597   &   1.84632   &   10.0084   &   9.22575   &   6.49839\\
\hline
\end{tabular} 
\caption{$\chi^2$ of SBP of simulated scenes at 2 Myr in XZ plane in three different regions and for the whole cluster. The smallest value is the closest one to SBP of R136.} 
\label{table:k2sbp-xz}  
\end{table*}

\begin{table*}
\centering 
\begin{tabular}{|c c c c c c c |} 
\hline
 & & &$0.2 - 150 M_{\odot}$ & & & \\
\hline
 & &Non segregated & & & Segregated & \\
\hline
Binary:  &  0\% & 30\% & 60\% & 0\% & 30\% & 60\%\\
\hline
reg1:  &  0.134300  &   0.121093  &   0.109488   &  0.296658   &  0.288149  &   0.291632 \\    
reg2:   &    0.313673   &  0.345682  &   0.233337   &   3.53505   &   3.24678   &   3.14732     \\
reg3:  &  2.24570   &   2.98828   &   2.46877    &  9.67167  &    11.5166   &   11.6881      \\
total:    &  2.64560    &  3.44385   &   2.80842   &   12.8372   &   14.2383    &  14.1527      \\
\hline
 & & &$0.5 - 150 M_{\odot}$ & & & \\
\hline
 & &Non segregated & & & Segregated & \\
\hline
Binary:  &  0\% & 30\% & 60\% & 0\% & 30\% & 60\%\\  
\hline
reg1:  &  0.199056  &   0.277567  &   0.201168  &   0.364366  &   0.311924   &  0.352169\\
reg2:  & 0.268080   &  0.388363 &    0.334479   &   2.08545   &   2.05616  &    1.62055\\
reg3:  &  1.24983    &  1.71160   &   2.10933   &   8.94076   &   7.85162   &   7.39265\\
total:   &   1.68033   &   2.27788   &   2.64436    &  10.9022    &  9.82424  &    9.14485\\
\hline
\end{tabular} 
\caption{$\chi^2$ of SBP of simulated scenes at 2 Myr in XY plane in three different regions and for the whole cluster. The smallest value is the closest one to SBP of R136.} 
\label{table:k2sbp-xy} 
\end{table*}

\subsection{MF slopes} \label{sec:mfslopes}
After having created series of synthetic scenes, in the first step photometry on each image have to be done. 
For the photometry on HST/WFPC2 data and also on the synthetic images we used STARFINDER (\citet{starfinder}) to extract the sources using analytical prepared Point Spread Function (PSF) related to the F814W filter. For HST/WFPC2-F814W we extracted 2660 stars and for the synthetic images, we extracted from 1759 up to 3193 sources depending on the scene.
Using BC tables (explained in Section \ref{sec:compareobs}), we estimated the photometric mass of detected stars in each imaging data for different clusters.
Furthermore we computed the MF for each simulated cluster at the evolution time of 2 Myr as following: 

$\log(N)=a \log(\frac{M}{M_{\odot}}) + b$

Table \ref{table:mfslopes-xz} and \ref{table:mfslopes-xy}, for the scenes in XZ and XY planes respectively, show the MF slopes for three regions ($r < 4".5$, $4".5 < r  < 9"$, $9" < r  < 13.5"$) and also for the whole cluster at time of 2 Myr. MF slope of R136 in the same filter is as follow:

$r < 4".5 : a = -0.39\pm0.24$

$4".5 < r  < 9" : a= -1.18\pm0.21$

$9" < r  < 13.5" : a= -0.97\pm0.09$

For the whole FoV: $a = -0.89\pm0.14$

\begin{figure*}
\centering 
        \includegraphics[width=8.cm]{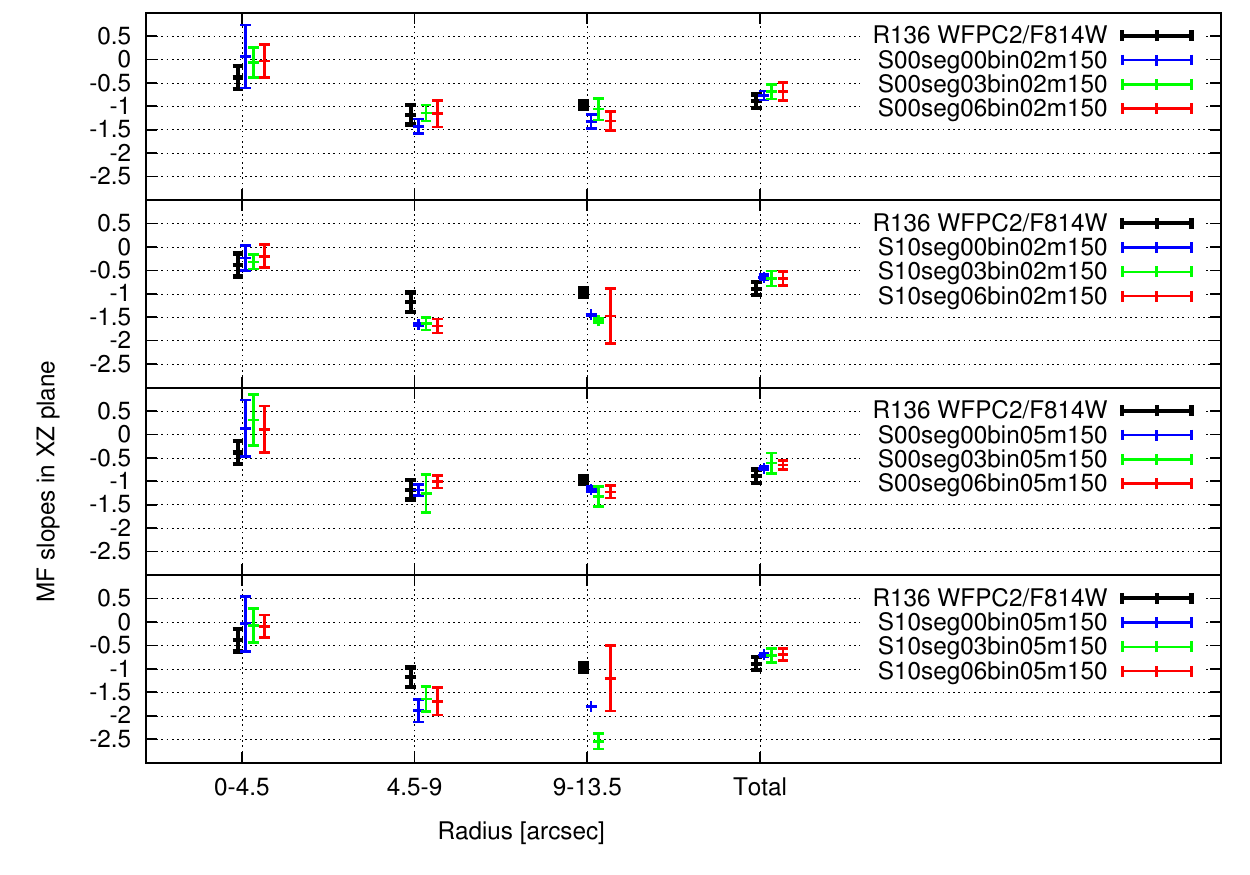}
        \includegraphics[width=8.cm]{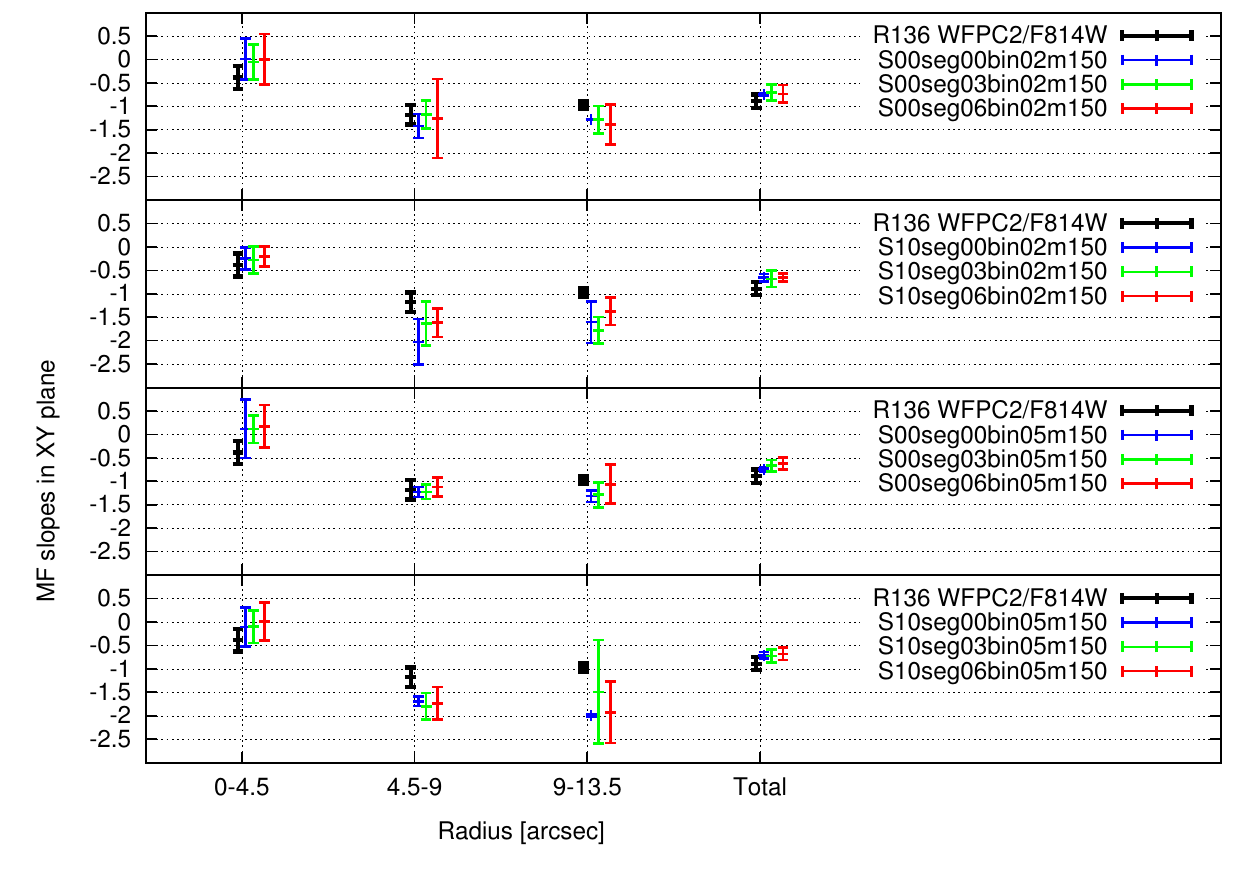}
\caption[]{MF Slopes from simulated synthetic scenes at 2 Myrs in XY (Right plots) and XZ (Left plots) plane. Black solid lines belong to R136 from HST/WFPC2 data taken in F814W filter. Blue, Green and Red represents clusters with 0\%, 30\% and 60\% initial binaries. 
\label{fig:mfplot}}
\end{figure*}

To better compare the slopes Figure \ref{fig:mfplot} on can compare them MF between simulations and R136 (Black solid line). In region 2 and 3 non-segregated simulated clusters exhibit closer values to the observed MF of R136.

\begin{table*}
\centering 
\begin{tabular}{|c c c c c c c |} 
\hline
 & & &$0.2 - 150 M_{\odot}$ & & & \\
\hline
 & &Not segregated & & & Segregated & \\
\hline
Binary:  &  0\% & 30\% & 60\% & 0\% & 30\% & 60\%\\
\hline
reg1:  &   $ 0.06\pm 0.67$  &   $-0.06\pm0.32$   &  $-0.03\pm0.35$  &  $-0.23\pm0.26$  & $-0.31\pm0.15$  &  $-0.19\pm0.24$    \\
reg2:  &   $-1.43\pm 0.15$ &   $-1.14\pm0.17$   &  $-1.16\pm0.28$  &  $-1.65\pm0.02$  & $-1.64\pm0.13$  &  $-1.68\pm0.15$    \\
reg3:  &   $-1.32\pm 0.14$ &   $-1.06\pm0.23$   &  $-1.31\pm0.20$  &  $-1.44\pm0.02$  & $-1.58\pm0.05$  &  $-1.47\pm0.58$    \\
total:  &   $-0.77\pm 0.09$ &   $-0.69\pm0.15$   &  $-0.68\pm0.19$  &  $-0.64\pm0.06$  & $-0.67\pm0.16$  &  $-0.67\pm0.67$    \\
\hline              
 & & &$0.5 - 150 M_{\odot}$ & & & \\
\hline
 & &Non-segregated & & & Segregated & \\
\hline
Binary:  &  0\% & 30\% & 60\% & 0\% & 30\% & 60\%\\ 
\hline
reg1:  &   $0.13\pm0.60$   &  $ 0.31\pm0.54$  &  $0.11\pm0.49$  &  $-0.04\pm0.59$  &  $-0.08\pm0.36$  &  $-0.09\pm0.24$ \\
reg2:  &  $-1.18\pm0.11$  & $-1.26\pm0.40$ &  $-1.01\pm0.14$ &  $-1.89\pm0.24$  &  $-1.64\pm0.26$  &  $-1.69\pm0.29$ \\
reg3:  &  $-1.17\pm0.06$  & $-1.33\pm0.22$ &  $-1.22\pm0.13$ &  $-1.79\pm0.01$  &  $-2.54\pm0.17$  &  $-1.20\pm0.69$ \\
total:  & $-0.72\pm0.04$   & $-0.61\pm0.22$ &  $-0.65\pm0.10$ &  $-0.70\pm0.04$  &  $-0.71\pm0.15$  &  $-0.70\pm0.12$ \\
\hline
\end{tabular} 
\caption{MF's slopes from simulated scenes at 2 Myr in XZ plane in three different regions and also for the the whole cluster. } 
\label{table:mfslopes-xz} 
\end{table*}

\begin{table*}
\centering 
\begin{tabular}{|c c c c c c c |} 
\hline
 & & &$0.2 - 150 M_{\odot}$ & & & \\
\hline
 & &Non-segregated & & & Segregated & \\
\hline
Binary:  &  0\% & 30\% & 60\% & 0\% & 30\% & 60\%\\ 
\hline     
reg1:  &  $  0.01\pm0.43 $ &  $-0.05\pm0.37 $  & $ 0.00\pm0.54$  &  $-0.24\pm0.23$  &  $-0.28\pm0.29$ & $-0.21\pm0.21$   \\
reg2:  &  $-1.42\pm0.26 $ &  $-1.18\pm0. 29$  & $-1.26\pm0.84$  & $-2.03\pm0.49$  &  $-1.64\pm0.47$ & $-1.61\pm0.30$    \\
reg3:  &  $-1.28\pm0.01 $ &  $-1.29\pm0.29 $  & $-1.38\pm0.43$  & $-1.60\pm0.44$  &  $-1.78\pm0.28$ & $-1.37\pm0.29$   \\
total:  &  $-0.75\pm0.03 $ &  $ -0.70\pm0.17$  & $-0.73\pm0.19$  & $-0.66\pm0.08$  &  $-0.68\pm0.17$ & $-0.65\pm0.07$    \\
\hline                        
 & & &$0.5 - 150 M_{\odot}$ & & & \\
\hline
 & &Non-segregated & & & Segregated & \\
\hline
Binary:  &  0\% & 30\% & 60\% & 0\% & 30\% & 60\%\\ 
\hline
reg1:  &  $0.12 \pm0.62$    &  $0.11\pm0.29$     & $0.17\pm0.46$    &  $-0.10\pm0.41$  & $-0.10\pm0.35$  &  $0.01\pm0.40$    \\
reg2:  &  $-1.22\pm0.11$   &  $-1.23\pm0.15$   & $-1.12\pm0.20$   &  $-1.69\pm0.10$  & $-1.80\pm0.28$  &  $-1.73\pm0.35$  \\
reg3:  &  $-1.32\pm0.12$   &  $-1.29\pm0.27$   & $-1.06\pm0.42$   &  $-1.99\pm0.02$  & $-1.49\pm1.11$  &  $-1.93\pm0.65$  \\
total:  &  $-0.75\pm0.05$   &  $-0.66\pm0.12$   & $-0.62\pm0.13$   &  $-0.71\pm0.07$  & $-0.72\pm0.14$  &  $-0.68\pm0.13$  \\
\hline
\end{tabular} 
\caption{MF's slopes from simulated scenes at 2 Myr in XY plane in three different regions and also for the the whole cluster. } 
\label{table:mfslopes-xy}  
\end{table*}

\subsection{Half-light radius}\label{sec:rhl}
In a next step, we estimated the half-light radius ($\rm{R}_{hl}$) of R136 versus those of synthetic scenes from the simulations at time 2 Myr. 

In order to compare the simulated scenes to the observed R136 we computed $\rm{R}_{hl}$ in the same FoV as to HST imaging data (32.5"x32.5" in terms of PC). Corresponding results are outlined in Table \ref{table:rh}. It can be seen that, unlike what $\rm{R}_{half-mass}$ shows in the simulations, non-segregated clusters have larger half light radii than segregated ones.

At time 2Myr, clusters with $M_{min} = 0.5 M_{\odot}$ have larger $\rm{R}_{hl}$ than clusters with $M_{min} = 0.2 M_{\odot}$. It means that clusters with larger number of low-mass (high-mass) stars have smaller (larger) half-light radius. 
Also $R_{hl}$ of the clusters with 0\% and 60\% initial binaries are more close together and both are larger than clusters with 30\% initial binaries.

\begin{table*}
\centering 
\begin{tabular}{|c |  c c c c c c |} 
\hline 
Model  & Seg & BF & $M_{min}$ & $M_{max}$ &   $R_{hl}$ (XY)&  $R_{hl}$(XZ) \\ [0.5ex]  
  &  &  & [$M_{\odot}]$ & [$M_{\odot}$] &   [pc]&  [pc] \\ [0.5ex]  
\hline
S00seg00bin05m150 & 0.0 & 0.0 &0.5 &150 &  0.43& 0.50\\
S10seg00bin05m150 & 1.0 & 0.0 &0.5 &150 &  0.30& 0.28\\

S00seg03bin05m150 & 0.0 & 0.3 &0.5 &150 & 0.40& 0.43\\
S10seg03bin05m150 & 1.0 & 0.3 &0.5 &150 & 0.26& 0.26\\

S00seg06bin05m150 & 0.0 & 0.6 &0.5 &150 & 0.46& 0.48\\
S10seg06bin05m150 & 1.0 & 0.6 &0.5 &150 & 0.29& 0.26\\

S00seg00bin02m150 & 0.0 & 0.0 &0.2 &150 & 0.42& 0.43\\
S10seg00bin02m150 & 1.0 & 0.0 &0.2 &150 & 0.27& 0.27\\

S00seg03bin02m150 & 0.0 & 0.3 &0.2 &150 & 0.36& 0.38\\
S10seg03bin02m150 & 1.0 & 0.3 &0.2 &150 & 0.23& 0.25\\

S00seg06bin02m150 & 0.0 & 0.6 &0.2 &150 & 0.42& 0.43\\
S10seg06bin02m150 & 1.0 & 0.6 &0.2 &150 & 0.25& 0.27\\
\hline
\end{tabular} 
\caption{$\rm{R}_{hl}$ calculated for different simulated scenes at 2Myr in XY and XZ plane.} 
\label{table:rh}
\end{table*}

\subsection{Neighbor radius}\label{sec:neighbor}

Neighbor radius ($R_{neighbor}$) is an arbitrary distance to find 100 neighbor stars from a given star. In the core of the cluster that has a higher stellar density, this radius is smaller than in the outer regions of the cluster. For R136 we calculated this radius for all the sources that have been detected in the HST image. We carried the same procedure on every simulated scene, with different characteristics, i.e. segregated or not, different binary fractions, etc... both in XY and XZ planes. Figure \ref{fig:r100xz} and \ref{fig:r100xy} depict the corresponding results. Red dots in each plot correspond to R136, Green and Blue dots correspond to not-segregated and segregated clusters. One can conclude that for all the simulated scenes, non-segregated clusters better fit to the R136 by HST considering the general trend of the slope.


\section{Discussion of the results}\label{sec:discussion}
We carried out a study of R136 in two steps. In a first  and as exhaustive as possible step based on the NBODY6 code,  we simulated a grid of synthetic young starburst compact clusters similar to R136, starting from its state-of-the-art basic parameters: i.e. age, distance, luminosity of individual member stars. In a second step we selected the likeliest synthesized images of R136 that match the observed visible wavelengths data from HST. The choice of this image provides a set of physical properties that explain best the expansion, the mass loss of the stellar populations in R136 as well as their binary fraction for instance.

Regarding the expansion of the cluster we found that in all cases segregated clusters expand more than non-segregated clusters, and that the former have larger $R_{h}$ than the latter. Clusters with stellar evolution (S-clusters) expand significantly around 3 Myr because of the evolution of very massive stars, that possess for some of them extreme winds of $10^{-8} - 10^{-5} [\rm{M}_{\odot}/\rm{yr}]$ (\citet{henny99}). 
Clusters with dynamical evolution (D-clusters) alone expand more due to the presence of more massive stars.

Regarding the mass loss of the cluster and escapers: the total mass loss of S-clusters is larger than for D-clusters. This is due to the evolution of massive stars themselves whilst D-clusters undergo escapers loss only. S-clusters containing more massive stars, present a larger mass loss in their early stages of evolution. D-clusters lose more escapers as the binary fraction increases.

On the other hand segregated clusters lose more binaries than non-segregated ones and in the special case of clusters with 30\% initial binaries made of low-mass stars, the binary loss is significantly larger.

Concerning periods and eccentricities: almost half of the low-mass binaries dissolve within 1 Myr. Clusters with stellar evolution lose their massive binaries according to the evolution of massive stars themselves. In all cases binaries keep the memory of initial eccentricities distribution during the first 4 Myr. For this period, only the massive binaries keep the memory of initial period distribution.

Having these general properties in mind we used synthetic scenes with the same observational resolution as HST/WFPC2-PC/F814W. For this purpose we computed a grid of bolometric correction tables which provide the flux of different stars in F814W filter using Geneva stellar evolution models and model atmosphere (TLUSTY for O,B stars and Kurucz for other spectral types). These synthetic images have the same pixel scale, spatial resolution and FoV of WFPC2 data on R136.

To conclude on the best  R136 from its synthesized images we used 4 different criteria based on: SBP, MF's slopes, $R_{hl}$ and $R_{conf}$.

We calculated the surface brightness profiles (SBP) of R136 and the synthetic scenes (in both XY and XZ planes). A $\chi^2$ criterion on the SBP permits to chose can the most probable synthesized R136 in three different regions and also for the whole cluster. SBP, $R_{hl}$ can easily be driven for each scene. Thus, using the photometry on each synthesized R136 image, we derived the mass of detected sources and plotted the mass function (MF) and $R_{neighbor}$.

Using the 4 criteria all together, even though they are not completely independent, we conclude that R136 is best represented by a non-segregated cluster (in $r<4\rm{pc}$). 

This result can explain the dominant mechanism for the formation of very massive stars among two main formation scenarii of gas accretion and collision between less massive stars that are explained in details by \citet{krumholz}. Initially a non-segregated cluster cannot provide sufficient dense conditions and a higher mean stellar mass in the core, for collisions to occur before the final stages of massive stars evolution. Presently,  no convincing evidence exists such as fragmentation, radiation pressure, photoionization and stellar winds to stop the growth of stars by accretion (\citet{krumholz}). So initially non-segregated clusters may better explain the formation of massive stars by accretion.

Effectively, accretion based models predict the low-mass companions, in addition to their massive companions, at separation of 100-1000 AU (\citet{kratter2006}, \citet{kratter2008}, \citet{kratter2010}, \citet{krumholz2012}) while the collisionally-formed stars will lack low-mass companions. These companions can be observationally detected by using high angular resolution and high contrast instruments like VLT/SPHERE (\citet{zurlo2014}), the future  E-ELT where R136 would be resolved as NGC3603 by the VLT and NGC3603 resolved by the VLT as the Trapezium cluster in Orion by a 1-2m class telescope.

Note that for the comparison of synthesized versus observed images of R136, we considered the median value of the extinction (in F814W filter) derived from 26 known O-type stars in the FoV of the HST data (Khorrami et al. in prep). If the spatial distribution of dust were taken into account according to the real position of massive stars in the cluster, the effect of confusion would even be worse. Since massive stars are expected to clear out dust from their neighborhood, the extinction would affect the low mass stars specially. This would make them undetectable in the visible wavelengths, that introduces an additional bias on the HST images reinforcing the segregation scenrio for R136 (\citet{ascenso2009}). 


\section{Conclusion}\label{sec:conclusion}
In this work we proposed a new approach to compare the results of the NBODY6 code with data from high contrast imaging observations obtained on large optical telescopes at their diffraction limit. In previous studies, the direct output of the code is compared with observational material where one compares for example the density profile of hundred thousands of stars with the density of a few thousands of them  extracted from observations in practice. Our method is rather based on synthesizing the observations directly from NBODY6 itself. These synthesized images are matched to real observations in a final step for their likeliness.

We based our study on data taken from HST/WFPC2 archives. For this we produced simulated scenes at the resolution of HST/WFPC2-pc/F814W images) and created BC tables which provide the flux of different stars in F814W filter using Geneva stellar evolution models and TLUSTY atmosphere model for O,B stars and Kurucz model for other spectral types. These synthetic images have the same pixel scale, spatial resolution and FoV as the WFPC2 data of R136 from HST. Note that our modeling of stellar members atmospheres could be improved by considering more appropriate atmosphere codes for the WR components of R136, for example using grids of model atmosphere like TLUSTY but including winds (\citet{Neugent15}). On the other hand our synthesized R136 clusters could be improved for their evolution by adding time-space depending gas potential to the model. This could ideally be included in NBODY6 code itself (private communication R. Wunsch).

In summary our study is in favor of the R136 to be a non-segregated cluster: a result contradicting the generally accepted picture. A result that deserves more exhaustive and systematic observations of R136 to be conclusive. Such observations should be carried out in as many spectral bands as possible: from the visible to optical and thermal IR wavelengths to overcome the confusion effect specially. This becomes possible using the VLT and high contrast AO imaging in the optical, future observations from space (JWST) or the E-ELT (\citet{Zinnecker2006}). Ultimately long baseline imaging interferometry from the ground should enable us to resolve the stellar binary components of R136 or similar compact clusters.

\begin{acknowledgements}
ZK is supported by the Erasmus Mundus Joint Doctorate Program by Grant Number 2012-1710 from the EACEA of the European Commission. We warmly thank Marcel Carbillet, Sambaran Banerjee, Andreas Kupper, Richard Wunsch for useful discussions and Fr\'ederic Th\'evenin for his careful reading of the original paper.
\end{acknowledgements}


\appendix
\section{Synthetic scenes}
\begin{figure*}
 \centering 
        \includegraphics[width=17.cm]{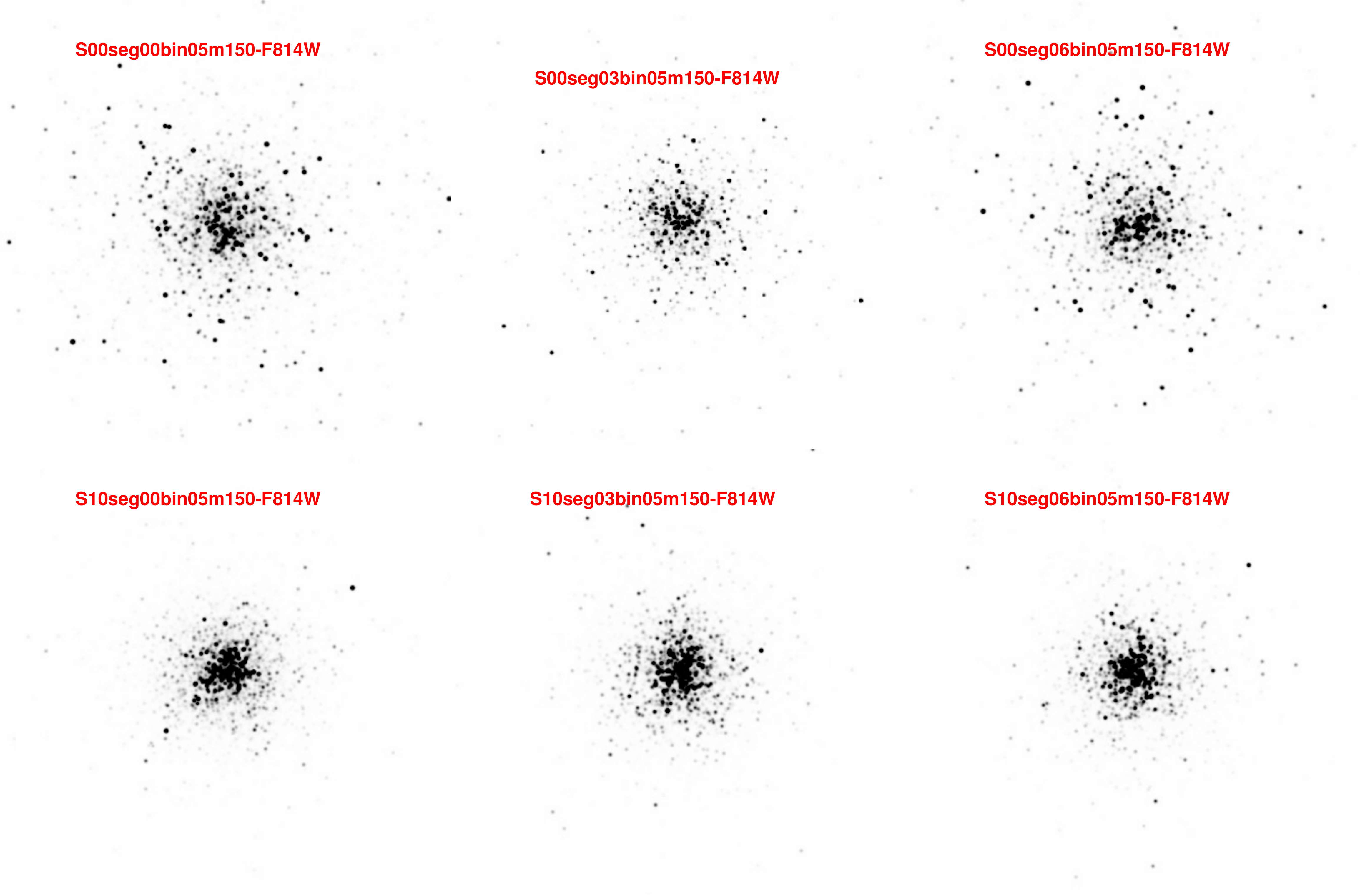}
\caption[]{Synthetic scenes from simulation of S-clusters with the mass range of $(0.5-150) \rm{M}_{\odot}$ at time 2 Myr in XZ plane. Not segregated: Upper images; Segregated clusters: Bottom images;  0\% initial binaries: left; 30\% initial binaries: middle; 60\% initial binaries: right
\label{fig:scenesxz05m150}}
\end{figure*}
\begin{figure*}
 \centering 
        \includegraphics[width=17.cm]{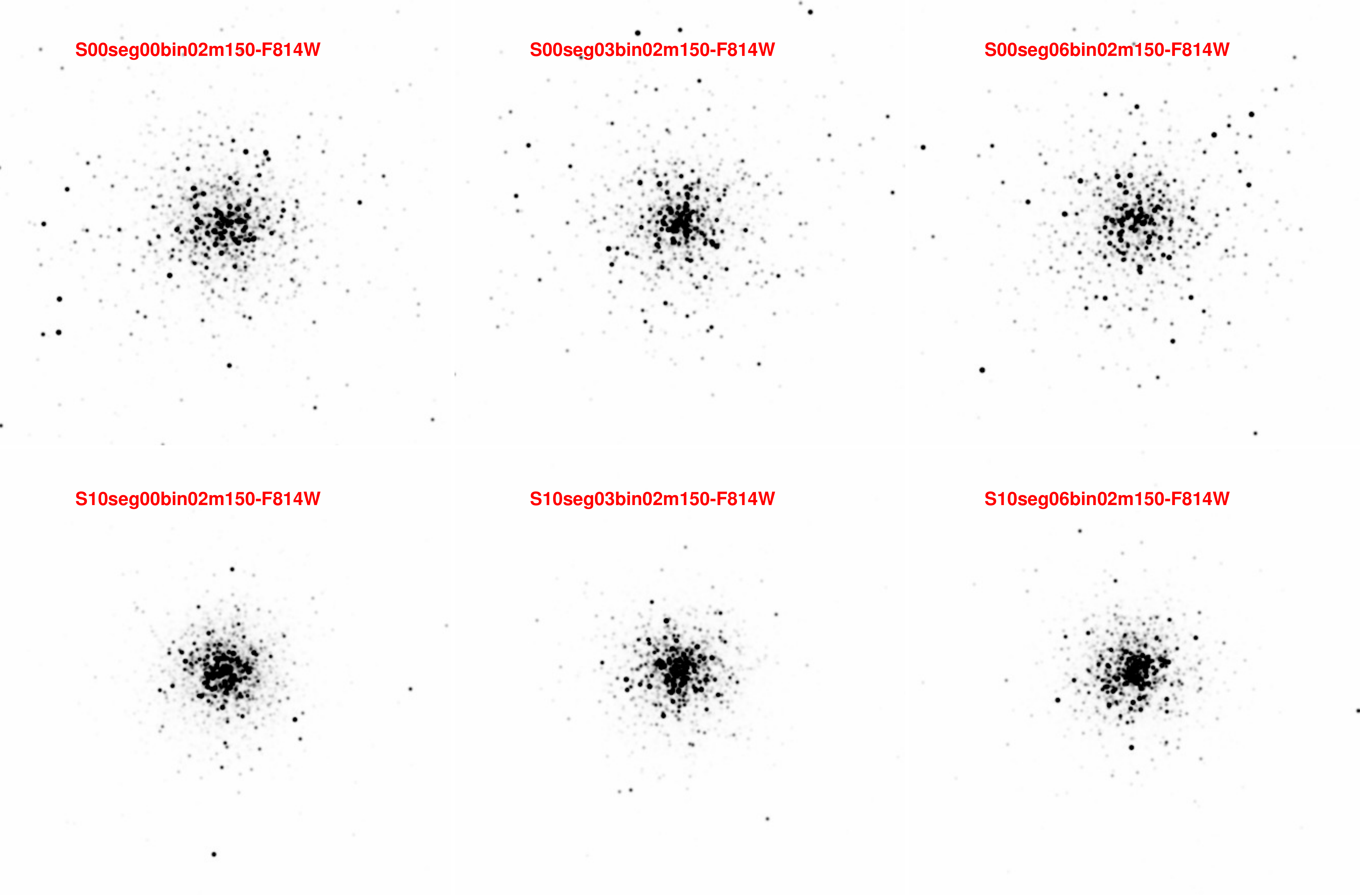}
\caption[]{Synthetic scenes from simulation of S-clusters with the mass range of $(0.2-150) \rm{M}_{\odot}$ at time 2 Myr in XZ plane. Not segregated: Upper images; Segregated clusters: Bottom images;  0\% initial binaries: left; 30\% initial binaries: middle; 60\% initial binaries: right
\label{fig:scenesxz02m150}}
\end{figure*}

\begin{figure*}
 \centering 
        \includegraphics[width=17.cm]{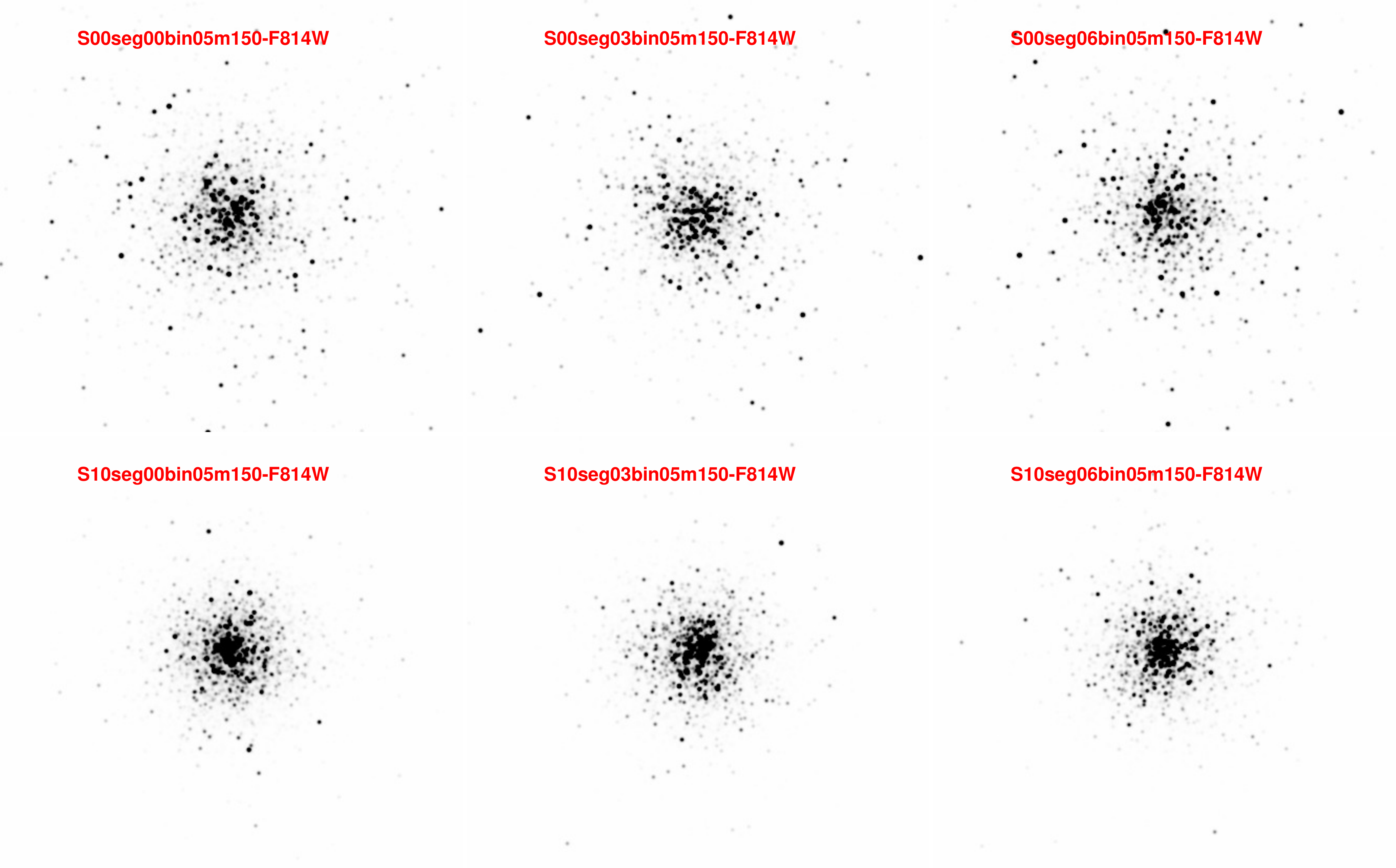}
\caption[]{Same as Figure \ref{fig:scenesxz05m150} but in XY plane.
\label{fig:scenesxy05m150}}
\end{figure*}

\begin{figure*}
 \centering 
        \includegraphics[width=17.cm]{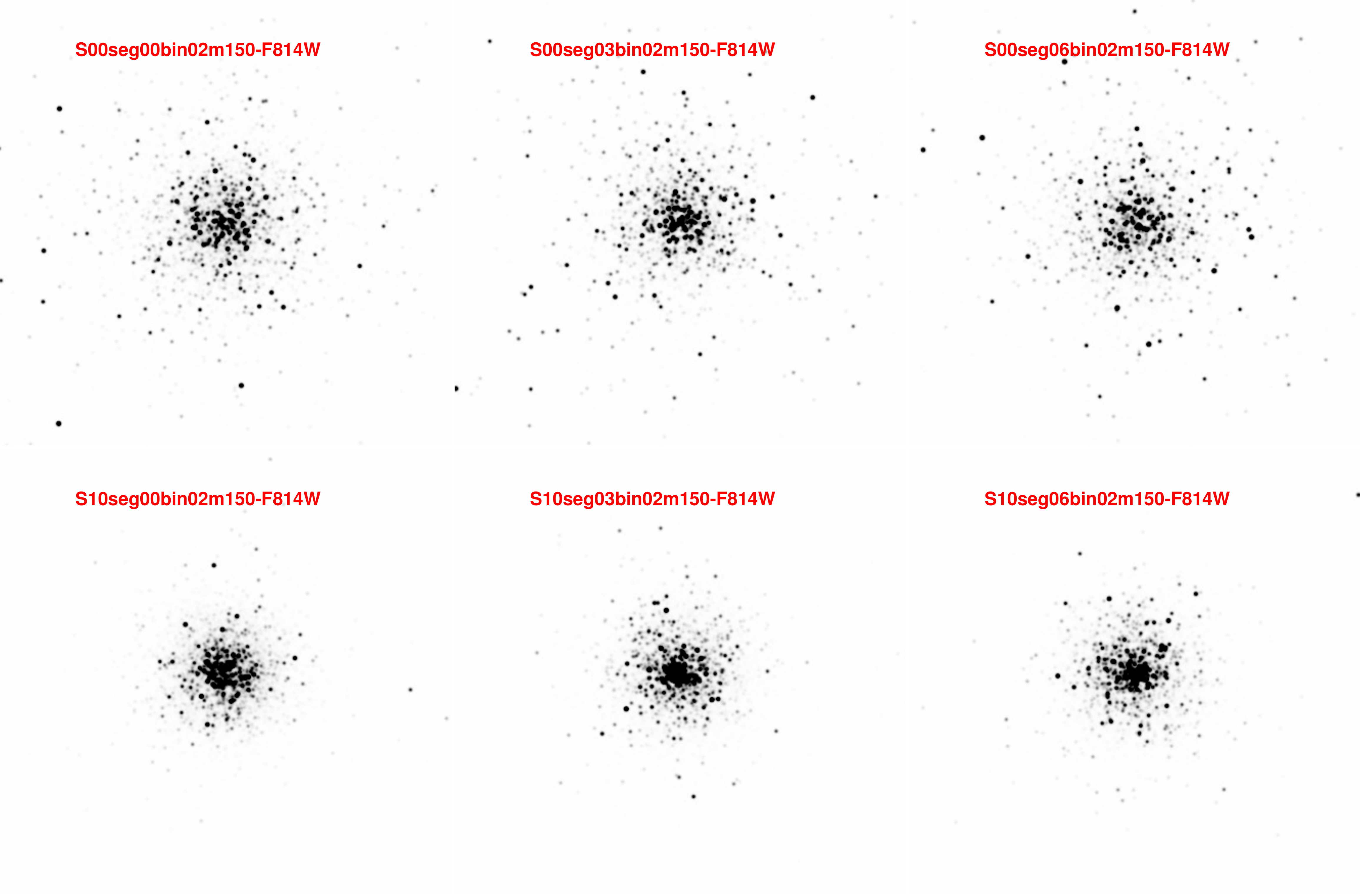}
\caption[]{Same as Figure \ref{fig:scenesxz02m150} but in XY plane.
\label{fig:scenesxy05m150}}
\end{figure*}

\section{Neighbor Radius}
\begin{figure*}
\centering 
\includegraphics[width=8.cm]{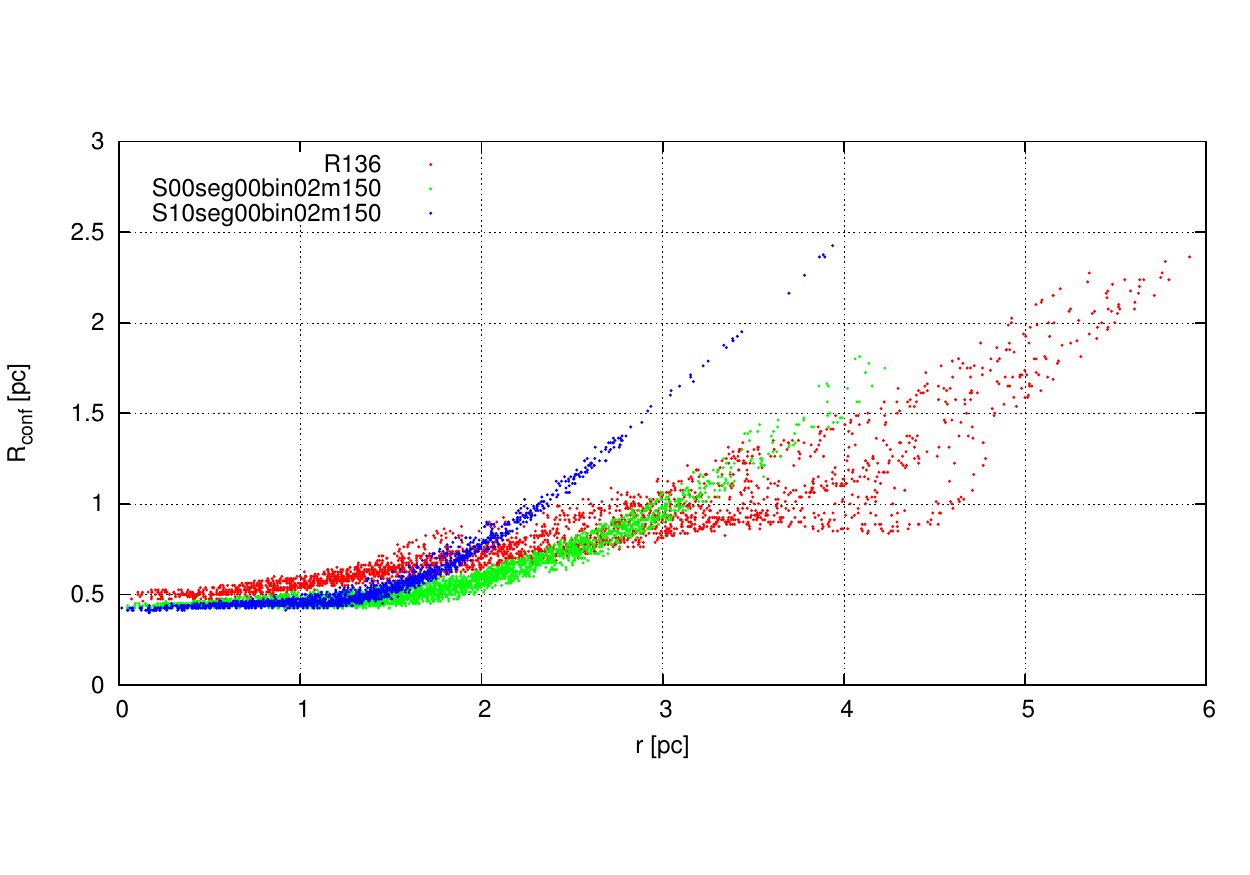}
\includegraphics[width=8.cm]{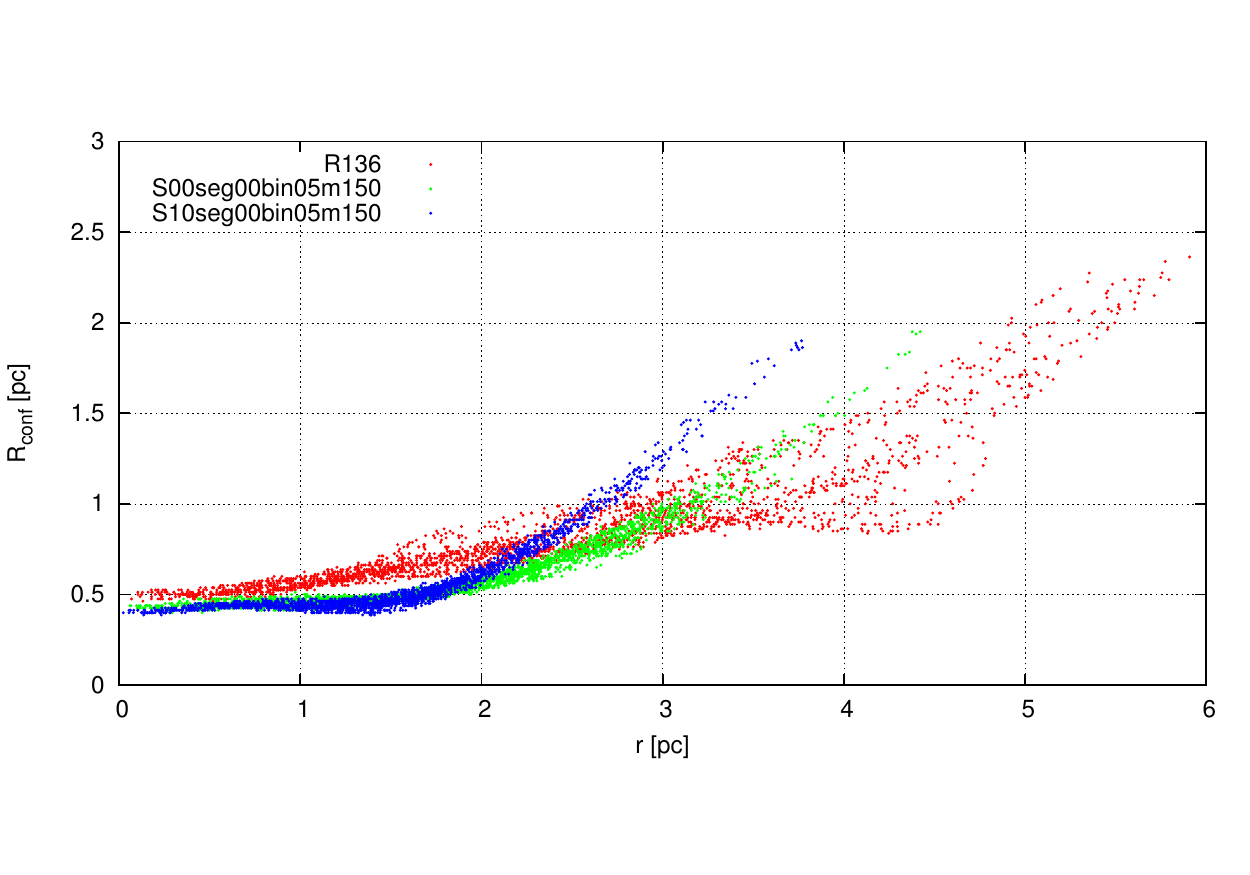}\\
\includegraphics[width=8.cm]{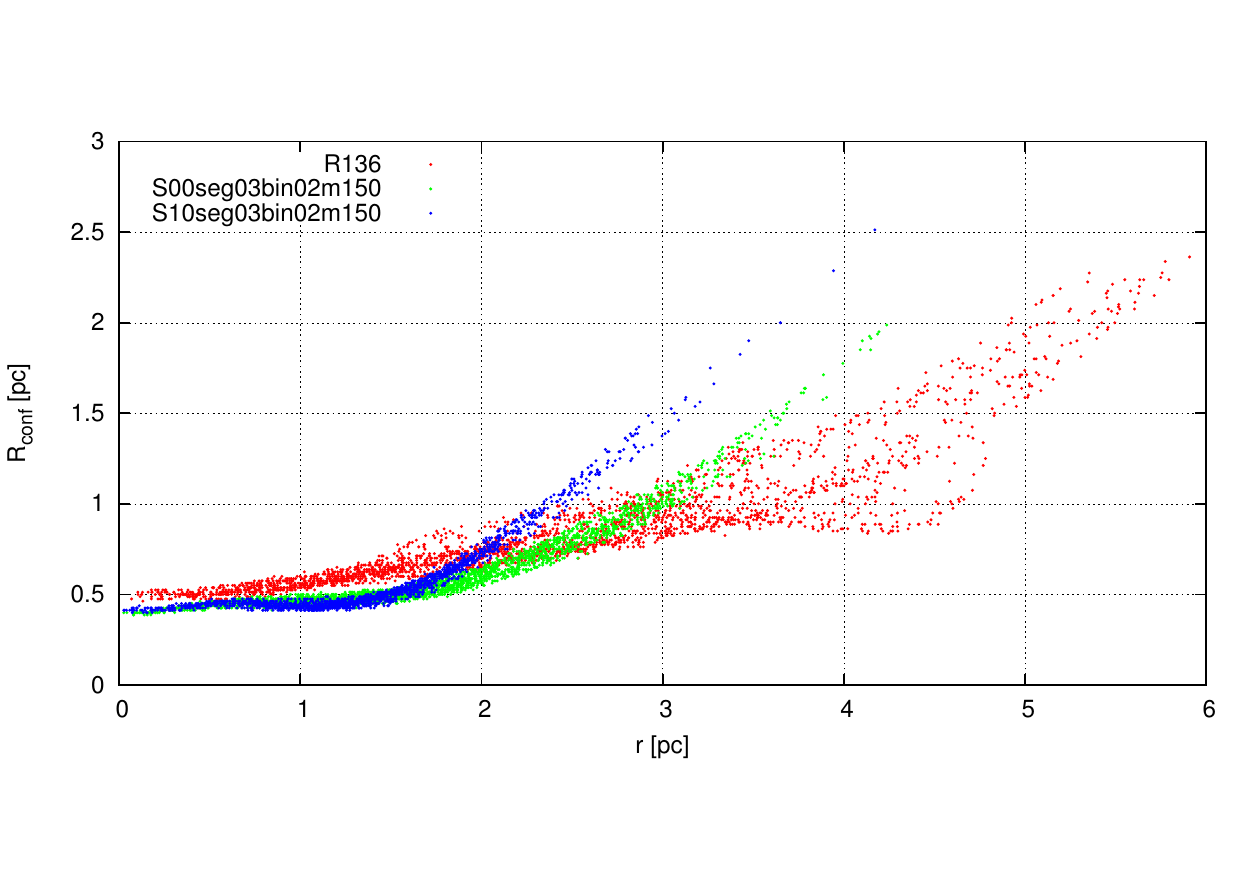}
\includegraphics[width=8.cm]{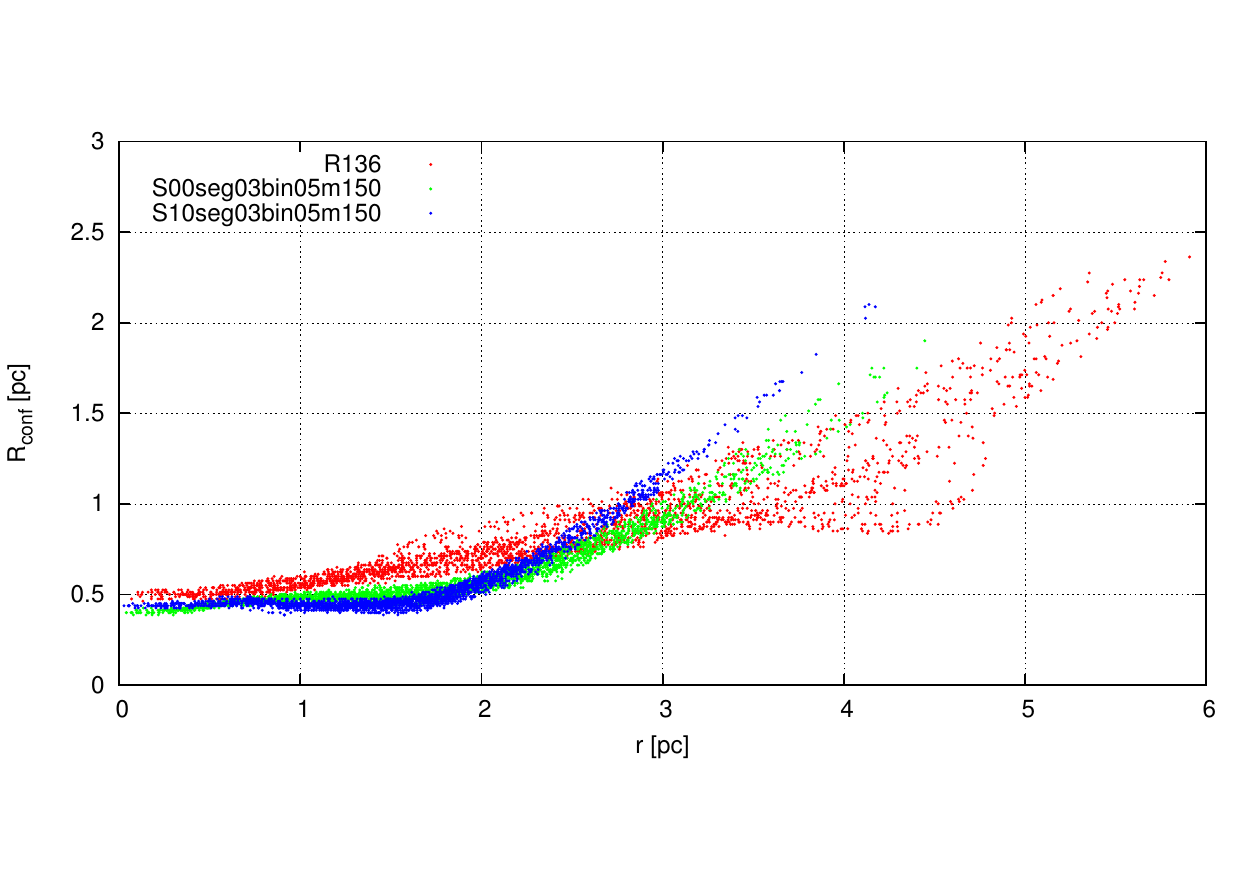}\\
\includegraphics[width=8.cm]{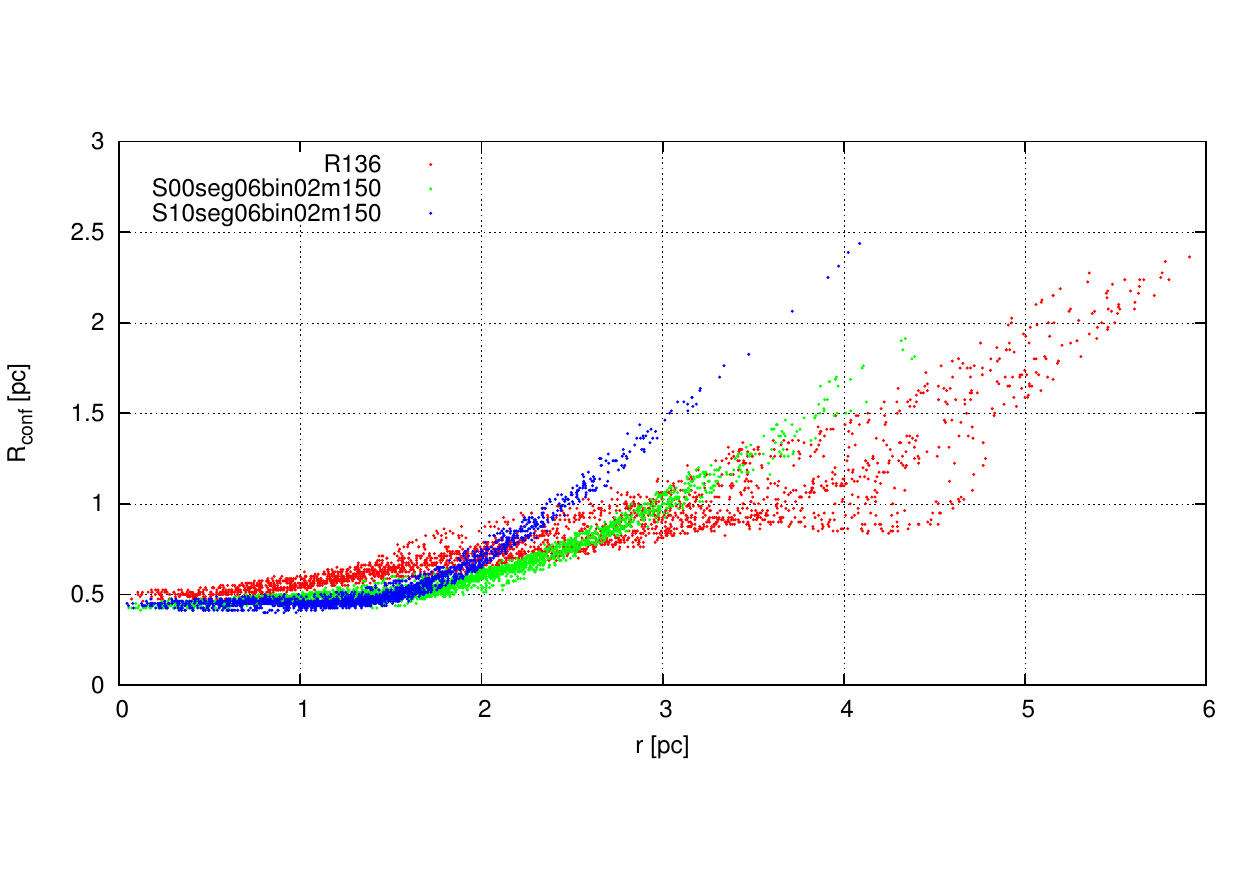}
\includegraphics[width=8.cm]{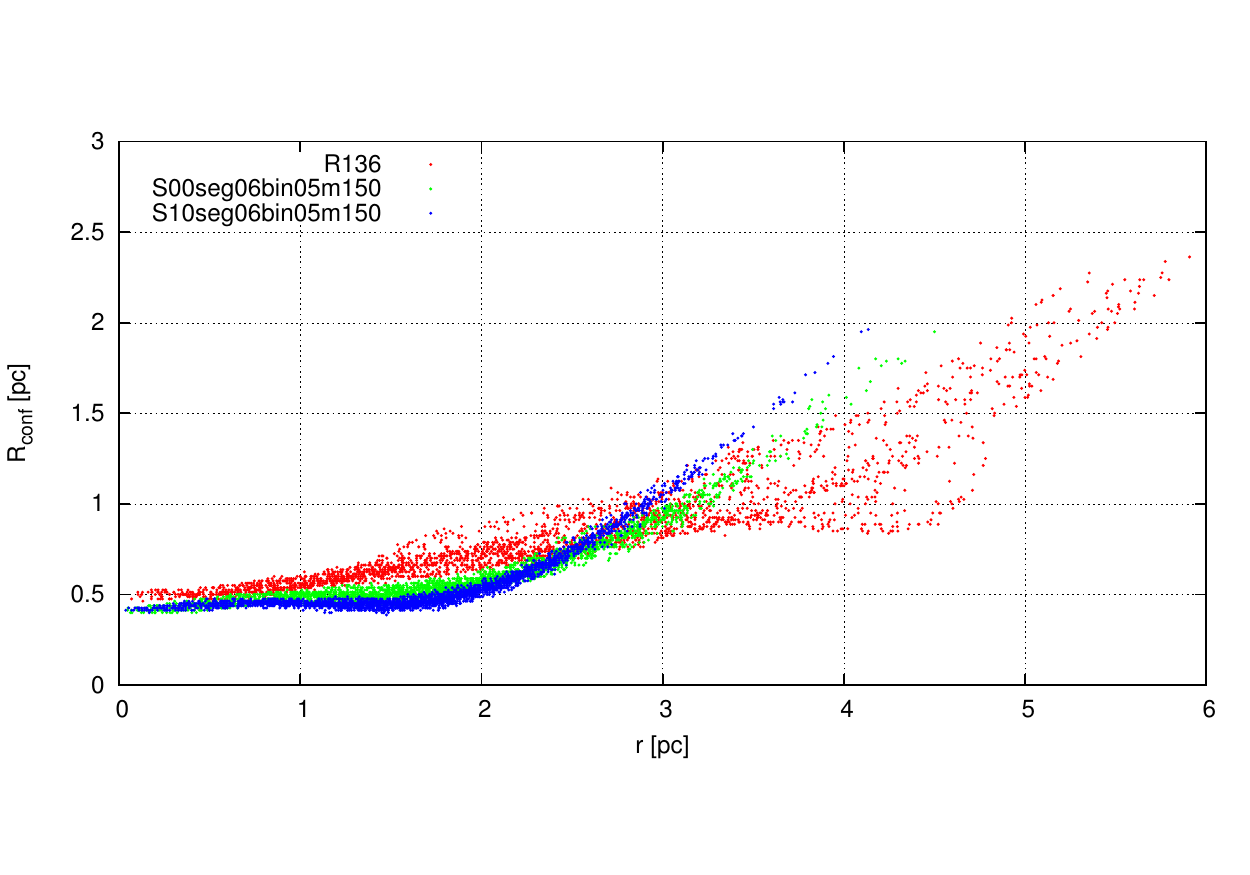}
\caption[]{$R_{neighbor}$ of simulated scenes in XZ-plane compare to R136 (Red dots). Blue and Green dots belong to the Segregated and Non-Segregated clusters. Left: clusters with the mass range of $(0.2 - 150) M_{\odot}$ and Right plots are for $(0.5 - 150) M_{\odot}$. Upper, middle and bottom plot represents clusters with 0\%, 30\% and 60\% initial binaries.
\label{fig:r100xz}}
\end{figure*}

\begin{figure*}
\centering 
\includegraphics[width=8.cm]{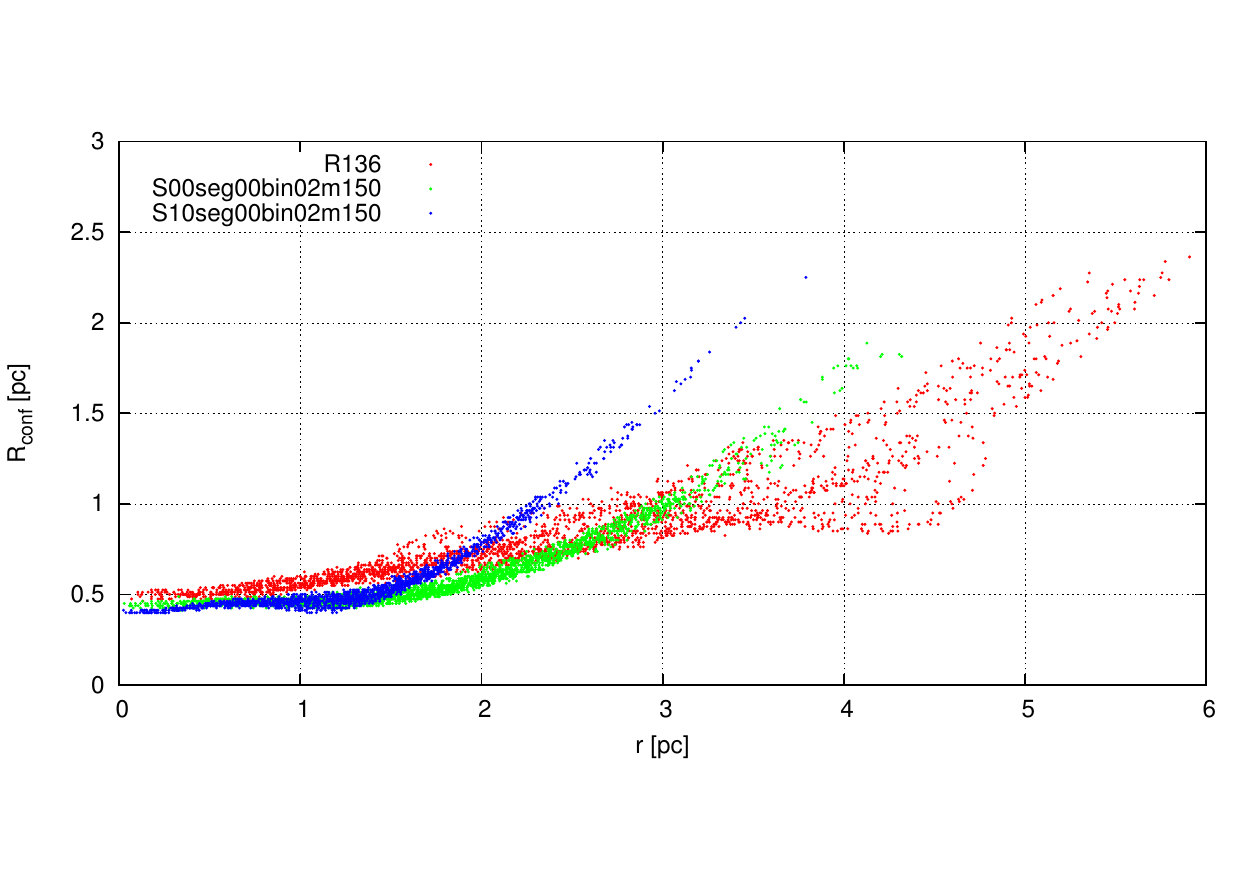}
\includegraphics[width=8.cm]{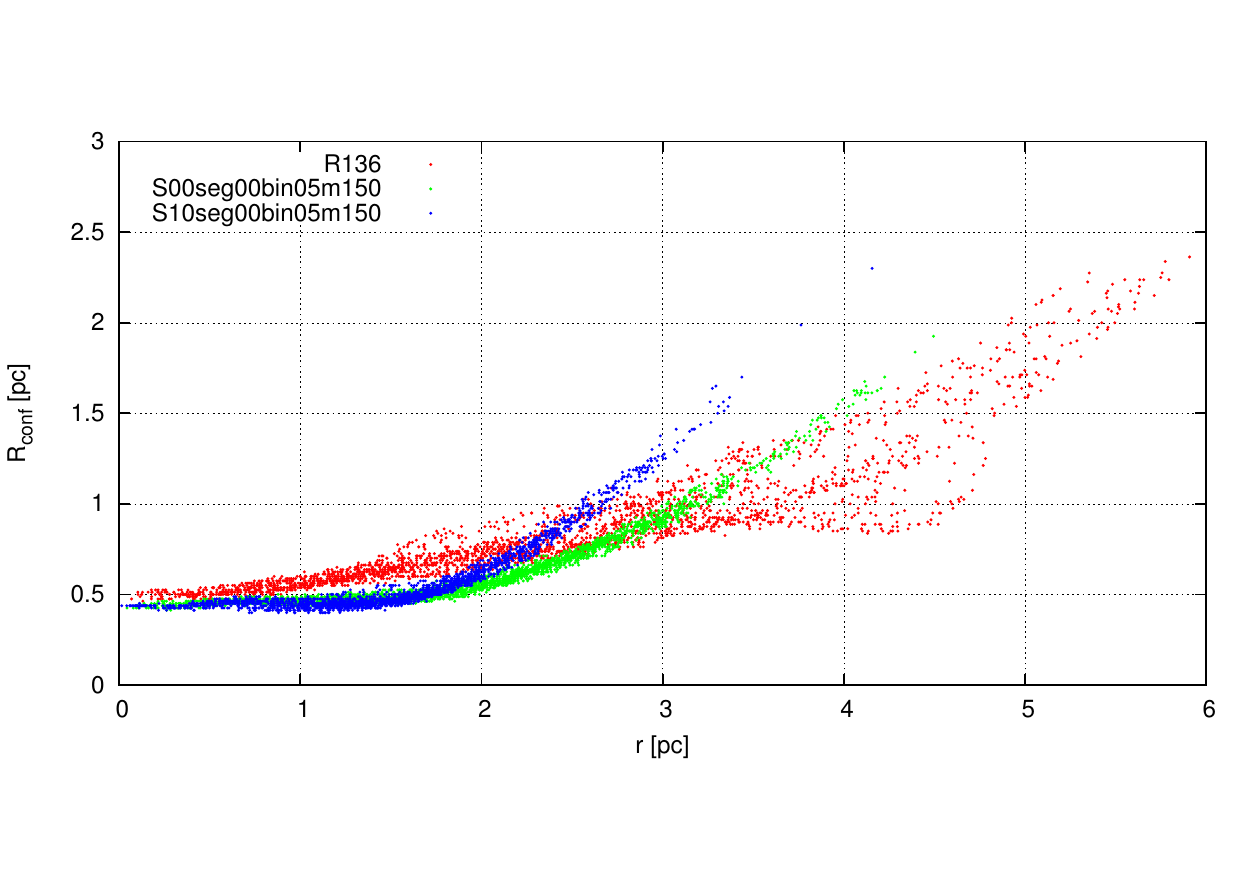}\\
\includegraphics[width=8.cm]{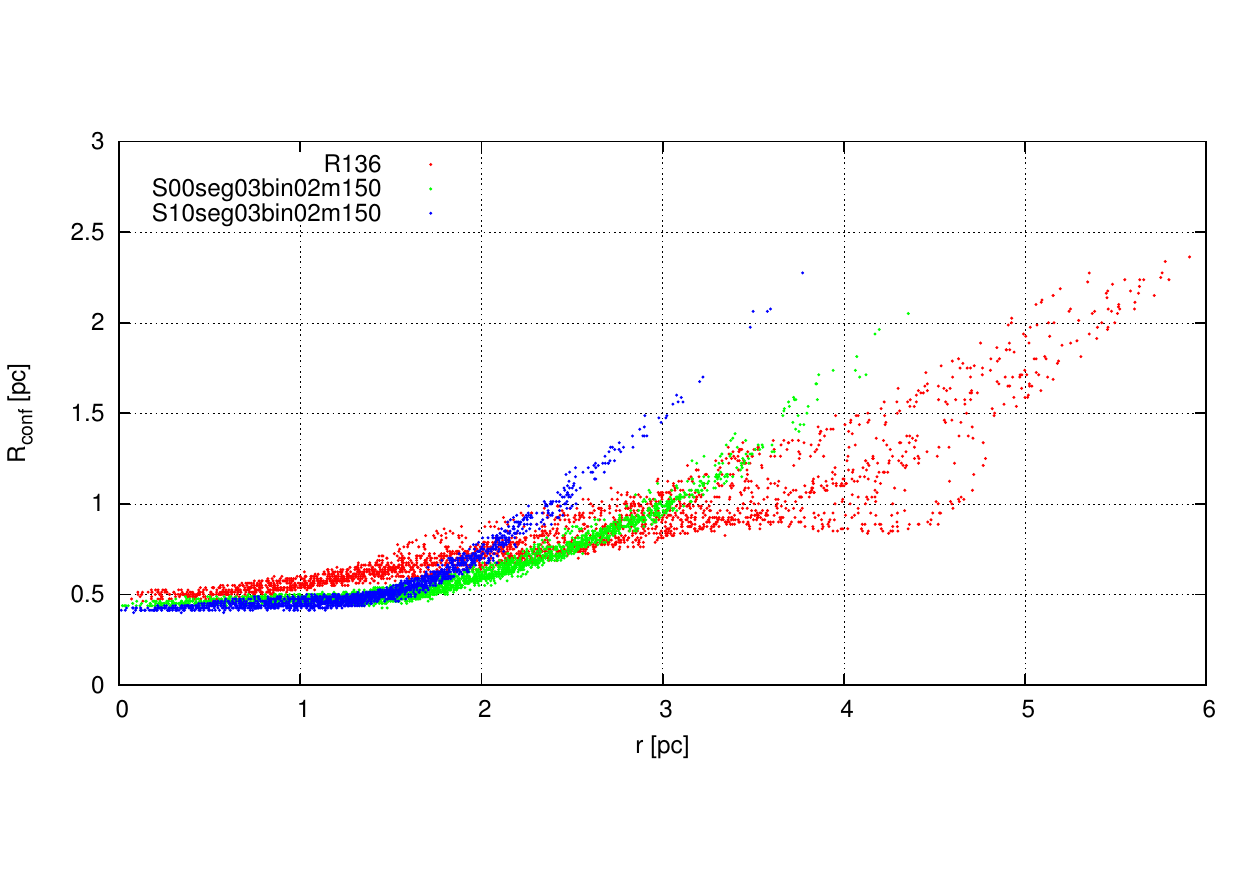}
\includegraphics[width=8.cm]{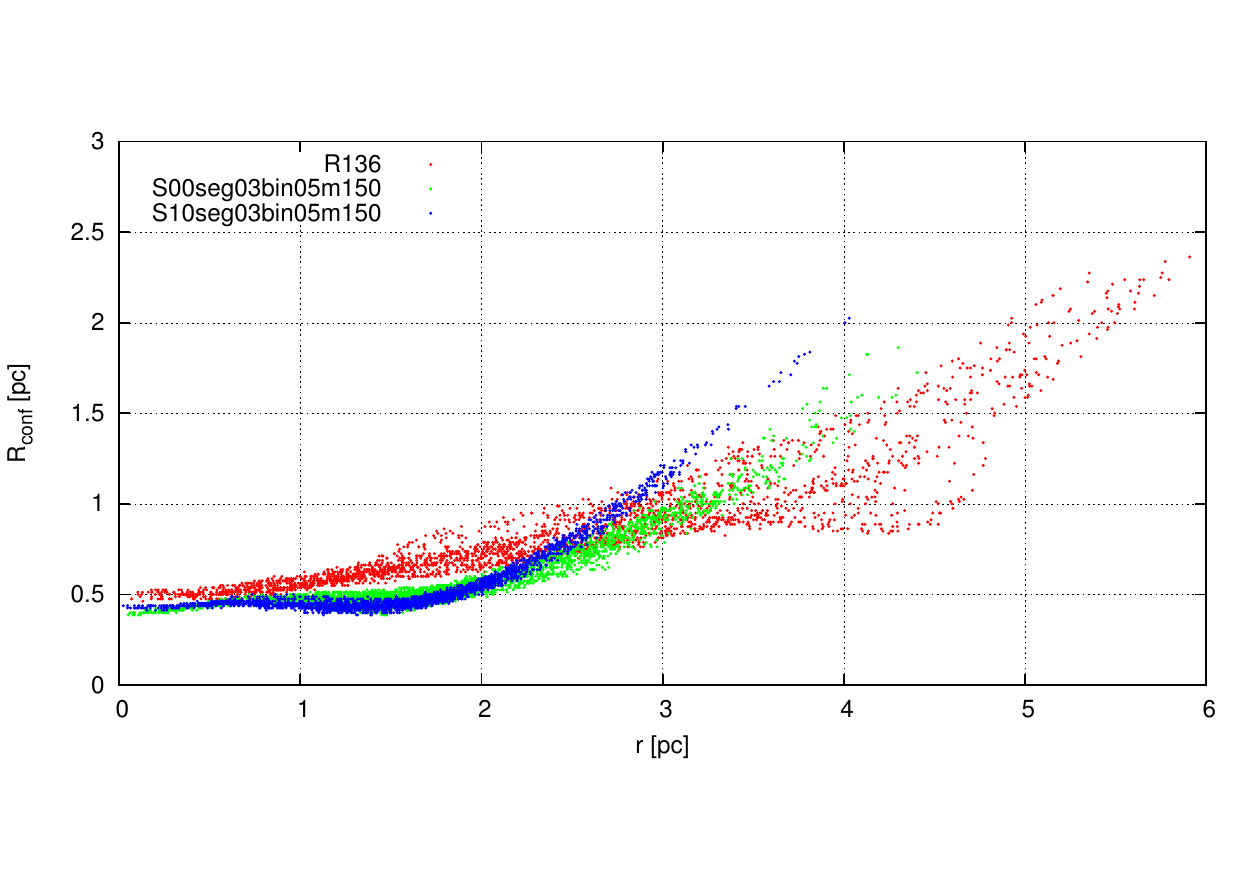}\\
\includegraphics[width=8.cm]{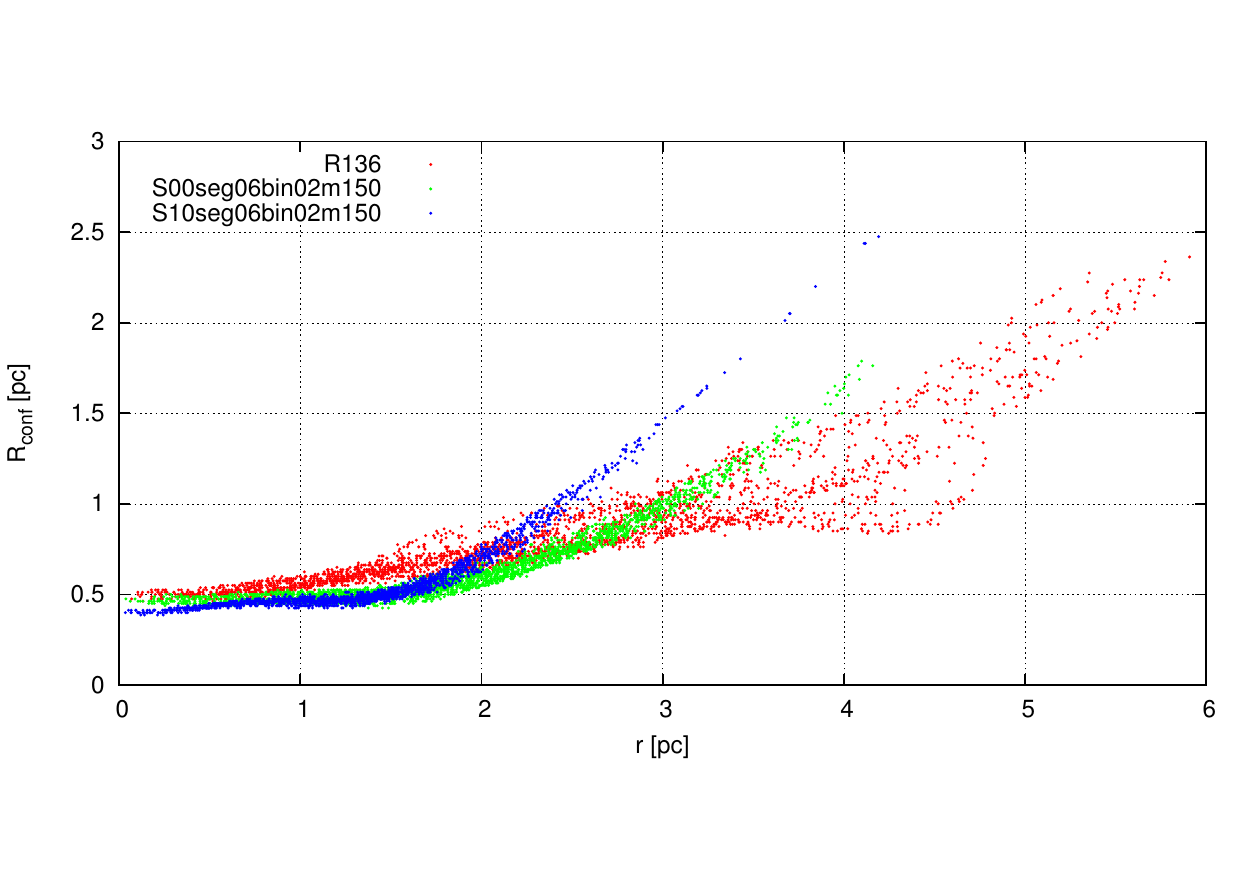}
\includegraphics[width=8.cm]{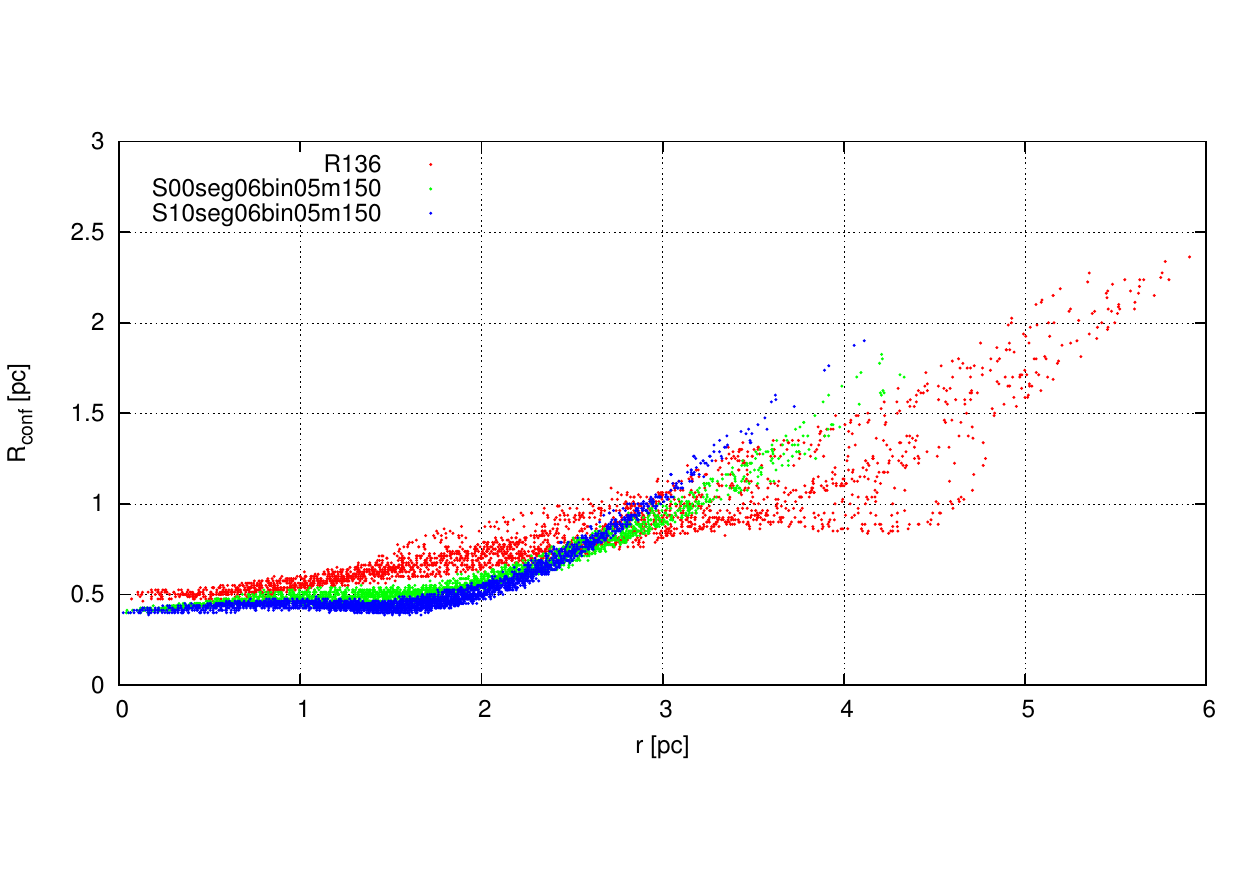}
\caption[]{$R_{neighbor}$ of simulated scenes in XY-plane compare to R136 (Red dots). Blue and Green dots belong to the Segregated and Non-Segregated clusters. Left: clusters with the mass range of $(0.2 - 150) M_{\odot}$ and Right plots are for $(0.5 - 150) M_{\odot}$. Upper, middle and bottom plot represents clusters with 0\%, 30\% and 60\% initial binaries.
\label{fig:r100xy}}
\end{figure*}

\section{Cumulative Surface brightness Profiles}\label{sec:cumSBP}

\begin{figure*}
\centering 
\includegraphics[width=16.cm]{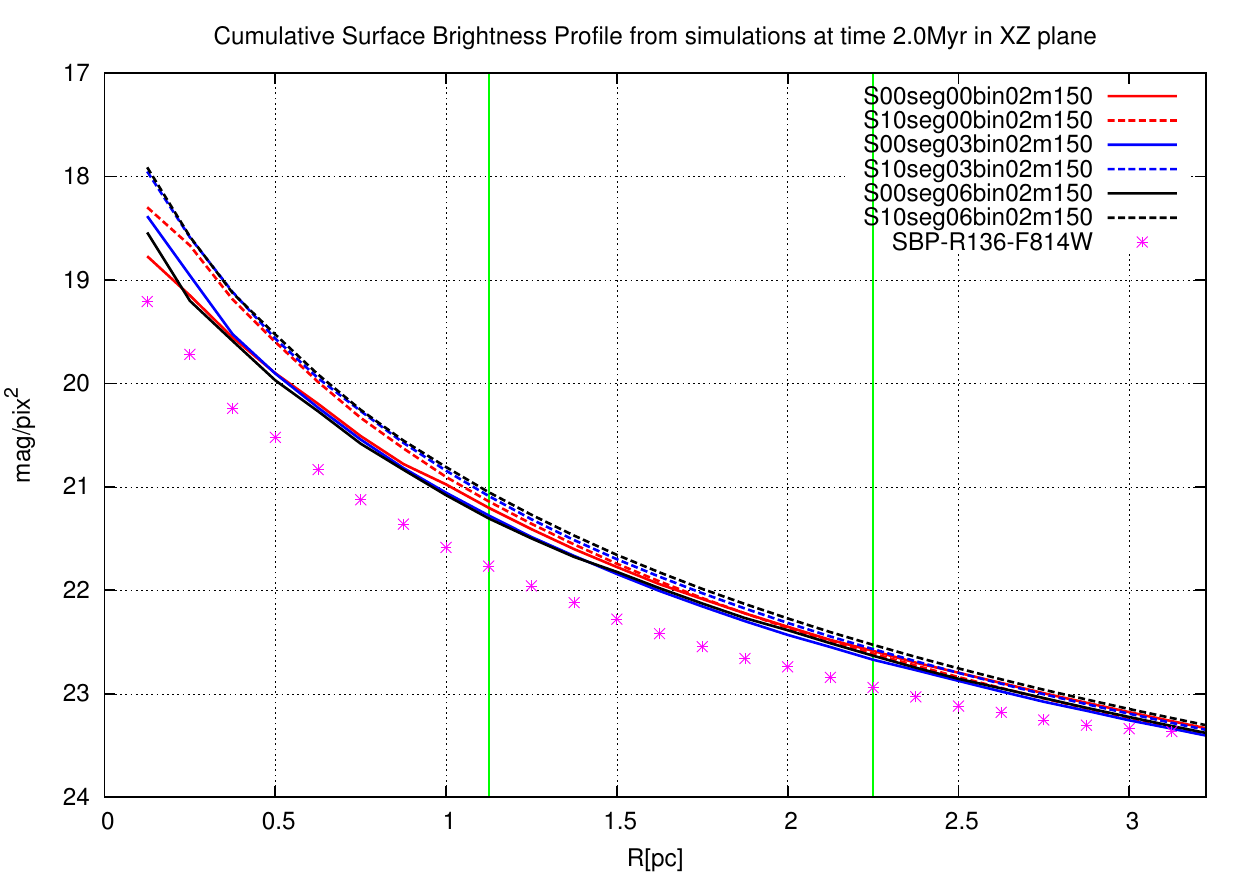}
\caption[]{Cumulative SBP for R136 (pink stars) and also for the synthetic simulations at time 2 Myr in XZ plan with the mass range of $(0.2 - 150) M_{\odot}$. Red, blue and black lines represent clusters with 0\%, 30\% and 60\% initial binaries. Solid and dashed lines belong to the Non-segregated and Segregated clusters. We fitted the function ($SBP(r)=r^a+constant$) to these data. The value of slope ($a$) for R136 and also for an example cluster with 30\% binary is:

R136: $a = 0.067 \pm 0.001$

S10seg03bin02m150 (Segregated) in three different regions: $a = 0.076 \pm 0.004$, $a = 0.094 \pm 0.001$, $a = 0.094 \pm 0.002$

S00seg03bin02m150 (Non-segregated) in three different regions: $a = 0.069 \pm 0.003$, $a = 0.097 \pm 0.002$, $a = 0.088 \pm 0.001$

In the core Non-segregated cluster has a slope similar to R136.
\label{fig:r100xz}}
\end{figure*}

\end{document}